\documentclass[fleqn,usenatbib]{rasti}

\usepackage{newtxtext,newtxmath}

\usepackage[T1]{fontenc}

\DeclareRobustCommand{\VAN}[3]{#2}
\let\VANthebibliography\thebibliography
\def\thebibliography{\DeclareRobustCommand{\VAN}[3]{##3}\VANthebibliography}

\usepackage{graphicx}	
\usepackage{amsmath}	

\title[Accretion flow and wind in TDEs]{Radiative hydrodynamical simulations of super-Eddington accretion flow in tidal disruption event: the accretion flow and wind}

\author[De-Fu Bu et al.]{
De-Fu Bu,$^{1}$\thanks{E-mail: dfbu@shao.ac.cn}
Erlin Qiao$^{2,3}$\thanks{E-mail: qiaoel@nao.cas.cn}
and Xiao-Hong Yang$^{4}$\thanks{E-mail:yangxh@cqu.edu.cn}
\\
$^{1}$Shanghai Astronomical Observatory, Chinese Academy of Sciences, 80 Nandan Road, Shanghai 200030, China \\
$^{2}$Key Laboratory of Space Astronomy and Technology, National Astronomical Observatory, \\ Chinese Academy of Sciences, Beijing 100012, China\\ 
$^{3}$School of Astronomy and Space Sciences, University of Chinese Academy of Sciences, 19A Yuquan Road, Beijing 100049, China \\ 
$^4$Department of Physics, Chongqing University, Chongqing 400044, China
}

\date{Accepted XXX. Received YYY; in original form ZZZ}

\pubyear{2022}

\begin{document}
\label{firstpage}
\pagerange{\pageref{firstpage}--\pageref{lastpage}}
\maketitle

\begin{abstract}
One key question in tidal disruption events theory is that how much of the fallback debris can be accreted to the black hole. Based on radiative hydrodynamic simulations, we study this issue for efficiently `circularized' debris accretion flow. We find that for a black hole disrupting a solar type star, $15\%$ of the debris can be accreted for a  $10^7$ solar mass ($M_\odot$) black hole. While for a $10^6M_\odot$ black hole, the value is $43\%$. We find that wind can be launched in the super-Eddington accretion phase regardless of the black hole mass. The maximum velocity of wind can reach $0.7c$ (with $c$ being speed of light). The kinetic power of wind is well above $10^{44} {\rm erg \ s^{-1}}$. The results can be used to study the interaction of wind and the circumnuclear medium around quiescent super-massive black holes.
\end{abstract}




\section{Introduction}
In galaxies, stars can move towards the supermassive black hole at the galaxy center. If the pericenter of the orbit of a star is equal to or smaller than the tidal disruption radius $R_{\rm T}$ \citep{Hills1975}, the star can be tidally disrupted, triggering the so-called tidal disruption events (TDEs, \cite{Rees1988}; \cite{Evans&Kochanek1989}). Roughly half of the stellar debris is unbound and can escape. The other bound half of the stellar debris falls back. The predicted fallback rate declines with time roughly as $\dot M_{\rm fb} \propto t^{-5/3}$.

The TDEs were first detected in the soft X-ray bands by the \emph{ROSAT} X-ray all-sky survey (see \cite{Komossa2015} for review). For these TDEs, the decline of their X-ray light curve is well consistent with the predicted $t^{-5/3}$ law. The X-ray is generated in the black hole accretion process. Thus, the consistency requires that the fallback rate roughly equals to the black hole accretion rate. This requirement is not obviously satisfied (\cite{Guillochon2015}; \cite{Shiokawa2015}). The stellar debris falls back to the orbit pericenter, which is much larger than the black hole horizon radius. In order to be accreted to the black hole, a viscous torque is required to transfer the angular momentum of the fallback debris. The fallback debris supplies gas to the viscous accretion flow. It is not guaranteed that all the fallback debris can be transported to the black hole horizon by the viscous accretion flow. From the theoretical point of view, it is quite necessary to study whether and how the black hole accretion rate correlates with the debris fallback rate. This is important to understand the X-ray light curve of TDEs.

For the optical/UV selected TDEs, the main puzzle is its origin of the optical/UV emission (see \cite{Velzen2020} and \cite{Gezari2021} for review). The inferred location of the optical/UV radiation is $\sim 10^{14-16}$ cm (\cite{Hung2017}; \cite{Velzen2020}; \cite{Gezari2021}). However, the accretion flow is predicted to have a size of several times of $10^{13}$ cm if one assumes that a solar type star is disrupted by a black hole with $10^6-10^7 M_\odot$. One proposed scenario is that the optical/UV emission is generated in the fallback debris colliding induced shock process (\cite{Piran2015}; \cite{Jiang2016}; \cite{Steinberg2022}). The location of shock is consistent with the observation inferred optical/UV radiation location. In the alternative `reprocessing' scenario, the soft X-ray/EUV emission generated very close to the black hole is reprocessed into optical/NUV bands by an surrounding optically thick and geometrically vertically extended envelope (\cite{Liu2021} \cite{Liu2017}; \cite{Metzger2017}; \cite{Metzger2022}; \cite{Wevers2022}; \cite{Loeb1997}; \cite{Coughlin2014}; \cite{Roth2016}) or wind (\cite{Strubbe2009}; \cite{Lodato2011}; \cite{Metzger2016}; \cite{Piro2020}; \cite{Uno2020}; \cite{Bu2022}; \cite{Parkinson2022}; \cite{Mageshwaran2023}). Recently, the presence of TDE winds has been directly confirmed by the UV and X-ray spectra (\cite{Yang2017}; \cite{Kara2018}; \cite{Parkinson2020}).

In analytical wind `reproessing' model, the properties of winds are arbitrary given \citep{Strubbe2009} due to the lack of the knowledge of the properties of TDEs wind. Efforts have been made to explore the properties of TDEs wind. By assuming that circularization is efficient, the properties of winds from a `circularized' super-Eddington accretion flow have been investigated by numerical simulation works (\cite{Dai2018}; \cite{Curd2019}). Howver, we note that these works just give the properties of winds at a snapshot around peak fallback rate. The time-evolution of winds is not studied. \cite{Thomsen2022} perform several discrete simulations with different accretion rate to study the time-evolution of TDEs wind. For this method, the winds from a early stage of the TDEs accretion flow have nothing to do with those from a later stage of the flow. To what extent this method can represent the real time-evolution of wind is not clear. \cite{Curd2023} studied the circularized accretion flow in TDEs. In order to be consistent with the situation of TDEs, the fallback debris is injected into the computational domain with a injection rate declining as $(t/t_{\rm fb})^{-5/3}$ law ($t_{\rm fb}$ is the debris fallback timescale). However, we note that in \cite{Curd2023}, a unrealistic shorter fallback timescale is employed in order to shorten the simulation time.  There are also works studying winds from the shock process (\cite{Jiang2016}; \cite{Lu2020}).

In addition to optical/UV emission, the TDEs wind may also be responsible for radio emission of TDEs (see \cite{Alexander2020} for review). The winds from TDEs can interact with the circumnuclear medium (CNM, \cite{Barniol2013}; \cite{Matsumoto2021}) or the dense clouds surrounding the central black hole (\cite{Mou2022}; \cite{Bu2023}), which can result in the formation of shocks. The power-law electrons which are responsible for radio emission can be accelerated in the shock process. The shock models are used to constrain the properties of winds, such as the velocity of wind.

Despite the importance of winds in understanding the electromagnetic radiation of TDEs, the detailed properties of TDEs winds are still poorly known. Although, there are many analytical and simulation works focusing on the winds from active galactic nuclei (AGN), the results can not be directly applied to TDEs. The reason is that the accretion flow in TDEs is quite different from that of an AGN. For example, the size of the accretion flow in TDEs is quite smaller than that of an AGN. Also, the accretion flow in TDEs has no quasi-steady state due to the fact that the gas supply rate to the flow declines as $t^{-5/3}$ law.

In order to study winds in TDEs, one need to take into account the specific conditions for TDEs. \cite{Bu2022} (hereafter BU22) performed hydrodynamic simulation with radiative transfer to study the `circularized' accretion flow in TDEs. In that paper, we take into account the specific conditions for TDEs. For example, we inject gas at 2 times of the pericenter of the orbit of the star, which is predicted to be the location of the accretion flow. The gas injection rate is set to declining as $(t/t_{\rm fb})^{-5/3}$ law to mimic the gas supply rate to the accretion flow due to the fallback of stellar debris. In BU22, we adopt the typical values for $t_{\rm fb}$, which is important for matching the special conditions for TDEs.

In this paper, based on the simulations in BU22, we study two important issues. The first one is the relationship between the black hole accretion rate and the fallback rate of the stellar debris. The second one is the property of the TDEs winds. The structure of this paper is as follows. In Section 2, we briefly introduce the numerical simulations of BU22. In section 3, we introduce the black hole accretion rate and properties of wind in TDEs. We summarize and discuss the results in Section 4.

\section{Numerical Simulations}

We briefly introduce the simulations in BU22. Two-dimensional axisymetric hydrodynamic simulations with radiative transfer are performed in BU22. The flux-limited diffusion approximation (\cite{Levermore1981}) is used to deal with radiation transfer.

The disrupted star is assumed to be solar type with radius $R_\ast=R_\odot$ and mass $M_\ast=M_\odot$, with $R_\odot$ being the solar radius. We consider the case in which the disrupted star moves on a parabolic trajectory towards the central black hole. We also assume that the orbital pericenter ($R_{\rm p}$) of the star is equal to the tidal disruption radius. Therefore, the penetration factor $\beta = R_{\rm T}/R_{\rm p} = 1$ . We simulate circularized accretion flow by assuming that when the disrupted stellar debris falls back, it can be very quickly circularized to form an accretion flow. Because of the angular momentum conservation, the circularized accretion flow forms at the circularization radius $R_{\rm C}$, which is two times the disruption radius $R_{\rm T}$. We use an anomalous stress tensor to mimic the angular momentum transfer by Maxwell stress.

We have two models. In model M7, the black hole mass $M_{\rm BH} = 10^7 M_\odot$. In model M6, we have $M_{\rm BH} = 10^6M_\odot$. The tidal radius for a $10^7M_\odot$ black hole is $R_{\rm T}=5R_s$ ($R_s$ is the Schwarzschild radius); while for a $10^6M_\odot$ black hole, $R_{\rm T}=47/2R_s$. We inject the circularized stellar debris around $R_C=2R_T$. The injection rate is set according to the theoretically predicted debris fallback rate $\dot M_{\rm inject} = \dot M_{\rm fb} = \frac{1}{3}(M_\ast/t_{\rm fb})(1+t/t_{\rm fb})^{-5/3}$, with $t_{\rm fb}$ being the debris fallback timescale. The fallback timescale $t_{\rm fb} \approx 40 \ {\rm days} \ (\frac{M_{\rm BH}}{10^6M_\odot})^{1/2} (\frac{M_\ast}{M_\odot})^{-1}(\frac{R_\ast}{R_\odot})^{3/2} $. Note that in our simulations, $t = 0$ corresponds to the one fallback timescale of the most bounded debris, at which the accretion begins rather the point at which the star is disrupted. We define the Eddington accretion rate as $\dot M_{\rm Edd} = 10 L_{\rm Edd}/ c^2$, with $L_{\rm Edd}$ being Eddington luminosity. The peak injection rate at $t=0$ for model M6 is $\sim 133 \dot M_{\rm Edd}$ and $\sim 4.7 \dot M_{\rm Edd}$ for model M7. The injected stellar debris is assumed to have a local Keplerian rotational velocity. The internal energy of the injected debris is assumed to be $1\%$ of the local gravitational energy. The simulations have computational domain in radial direction $2R_s \leq r \leq 10^5 R_s$ and in the $\theta$ direction $0 \leq \theta \leq \pi/2$. The resolution is $N_r \times N_\theta = 768 \times 128$.  Outflow boundary conditions are applied at the inner and outer radial boundary. At $\theta = 0$, we use the axisymmetric boundary conditions. At $\theta = \pi/2$, we use reflecting boundary conditons. The more details of the simulations are referred to BU22.

\section{Results}
The mass accretion rate onto the black hole is calculated at the inner radial boundary of the simulations ($2R_s$). Because we just simulate the region above the midplane, the accretion rate is 2 times the value above the midplane,
\begin{equation}
\dot M = 2 \times 2\pi (2R_s)^2 \int_0^{\pi/2} \rho \min(v_r,0) \sin \theta d\theta
\end{equation}
where, $\rho$ and $v_r$ are gas density and radial velocity, respectively.

We now introduce the method to calculate the mass flux of wind. Turbulence is present in our simulations. Thus, we can not judge the fluid element as wind only by $v_r > 0$. Because, the outward moving portion of a turbulence eddy also has positive velocity. As down by \cite{Curd2023}, we define gas which has positive Bernoulli parameter $Be > 0$ and $v_r > 0$ as wind. Following \cite{Curd2023}, before defining the Bernoulli parameter $Be$, we define the electron scattering optical depth first. Along a viewing angle $\theta$, the electron scattering optical depth is integrated from outer radial boundary inwards $\tau (\theta, r)=\int_{10^5R_s}^r \rho \kappa_{\rm es} dr'$. The electron scattering opacity $\kappa_{\rm es} = 0.34 {\rm cm}^2 {\rm g}^{-1} $. In the optically thick regions ($\tau_{\rm es} > 1$), the radiation is well coupled with gas and can contribute to acceleration of gas, so we treat it as contributing to the Bernoulli parameter. In the optically thin region, the radiation is decoupled from gas, so we do not include it to the calculation of the Bernoulli parameter. As down in \cite{Curd2023}, the Bernoulli parameter is calculated as follows,
\begin{align}
Be=\begin{cases}
\frac {1}{2} v^2 + \frac{\gamma_{\rm gas} e_{\rm gas}}{\rho} - \frac{GM_{\rm BH}}{r-R_s} + (1-\tau_{\rm es}^{-1/2})\frac{\gamma_{\rm rad} E_{\rm rad}}{\rho}      ~~~ (\tau_{\rm es} \geq 1) ~~ \\
\frac {1}{2} v^2 + \frac{\gamma_{\rm gas} e_{\rm gas}}{\rho} - \frac{GM_{\rm BH}}{r-R_s}     ~~~~~~~~~~~~~~~~~~~~~~~~~~~~~ (\tau_{\rm es} < 1)
\end{cases}
\end{align}
where $v$, $e_{\rm gas}$, $E_{\rm rad}$ and $G$ are gas velocity, gas internal energy density, radiation energy density and gravitational constant, respectively. We set specific heat ratio for gas $\gamma_{\rm gas} =5/3$; for radiation we set $\gamma_{\rm rad} = 4/3$. The calculation of Bernoulli parameter with optically thick radiation is also referred to \cite{Yoshioka2022}.

The mass flux of wind is calculated as follows,
\begin{equation}
\dot M_{\rm wind} = 2 \times 2\pi r^2 \int_0^{\pi/2} \max\left({\frac{Be}{|Be|}, 0}\right)  \rho \max(v_r,0) \sin \theta d\theta
\end{equation}
The kinetic power of wind is calculated as follows,
\begin{equation}
\dot E_{\rm wind} = 2 \times 2\pi r^2 \int_0^{\pi/2} \max\left({\frac{Be}{|Be|}, 0}\right)  \rho \frac{1}{2} v_r^2 \max(v_r,0) \sin \theta d\theta
\end{equation}

\begin{figure}
\begin{center}
\includegraphics[width=0.45\textwidth]{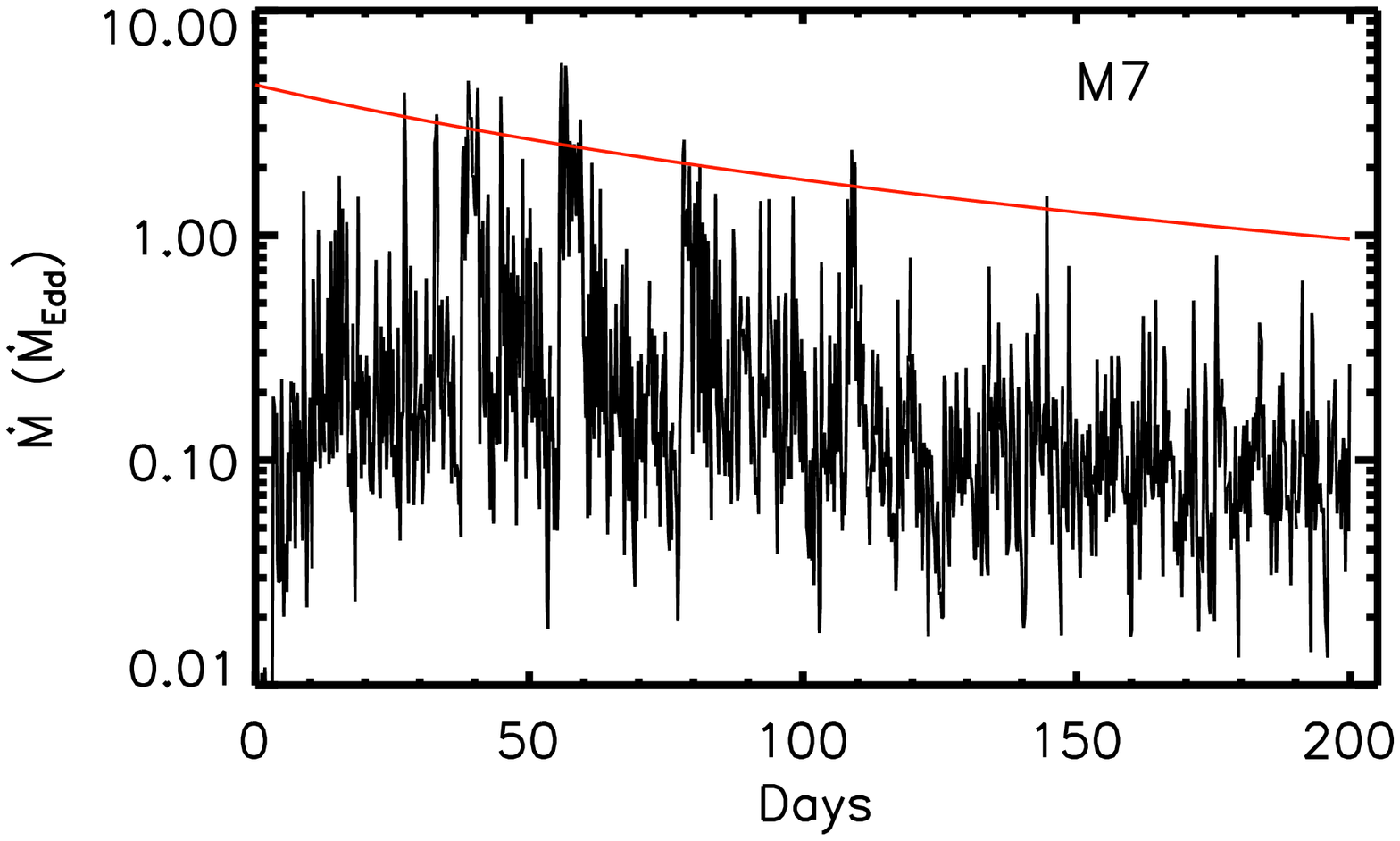}
\includegraphics[width=0.45\textwidth]{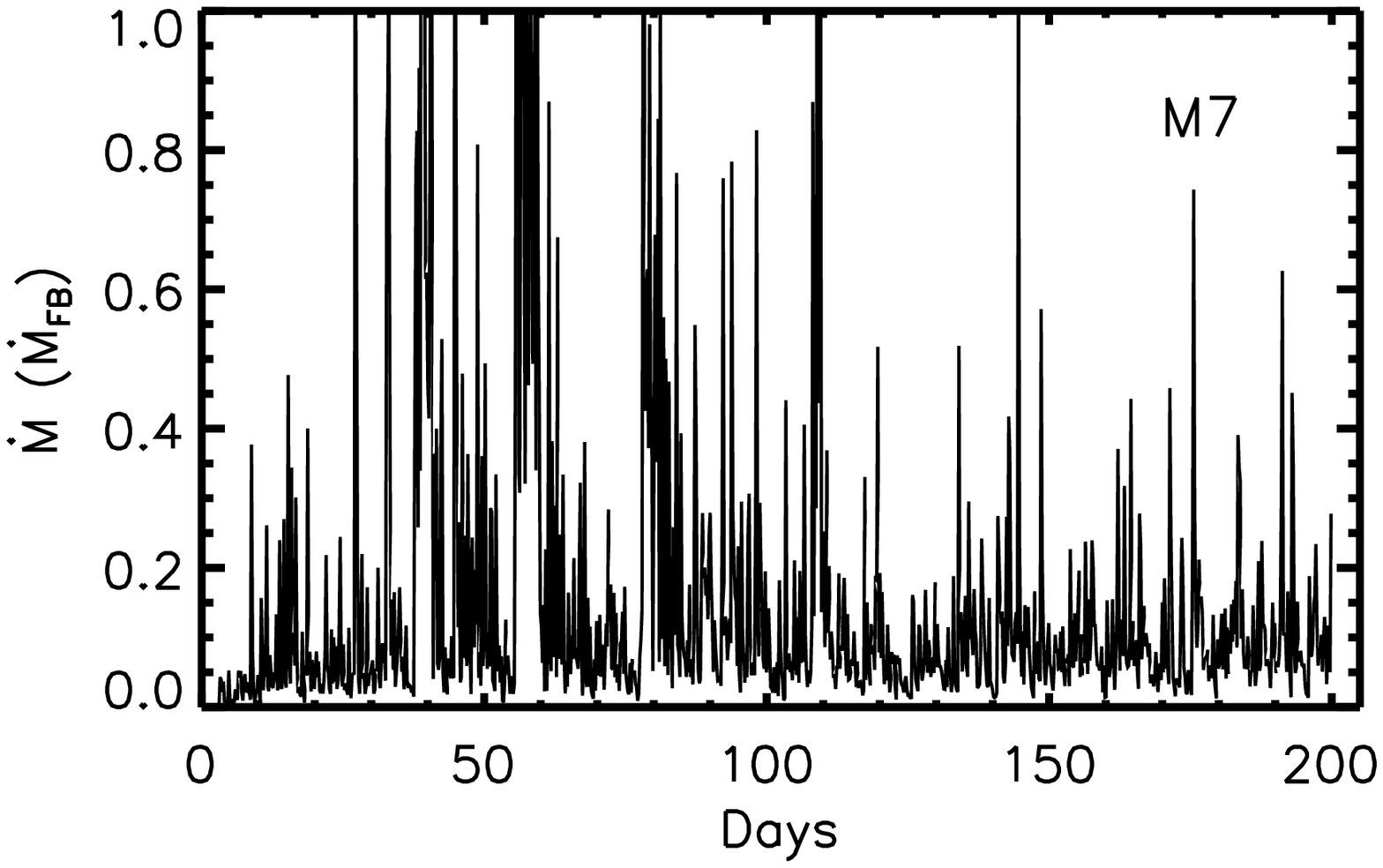}
\caption{Accretion rate for model M7. Top panel: time evolution of the black hole accretion rate (black line) and the stellar debris fallback rate (red line) in unit of Eddington accretion rate. Bottom panel: time evolution of the black hole accretion rate in unit of the stellar debris fallback rate.}
\label{fig:mdotM7}
\end{center}
\end{figure}

\subsection{Model M7}
In model M7, the central black hole mass is $10^7 M_\odot$. The circularized stellar debris is injected into the computational domain around 10$R_s$, which is 2 times the stellar orbital pericenter. In the presence of viscosity, an accretion flow forms, which transports gas towards the central black hole. The simulation covers 205 days since the peak fallback rate. The fallback rate is super-Eddington and the flow is radiation pressure dominated.

The top panel of Figure \ref{fig:mdotM7} shows the time evolution of the black hole accretion rate (black line) and the stellar debris fallback rate (red line) in unit of Eddington accretion rate. The bottom panel of Figure \ref{fig:mdotM7} shows the time evolution of the black hole accretion rate in unit of the stellar debris fallback rate. The black hole accretion rate is highly variable. The variability is due to the fact that the flow is quite turbulent. \cite{Yang2014} performed hydrodynamic radiation pressure dominated accretion flow. They also find that the flow is quite turbulent. Our result is consistent with that in \cite{Yang2014}. The turbulence is due to that the flow is convectively unstable. On average, the black hole accretion rate is significantly smaller than the stellar debris fallback rate (see the bottom panel). We quantitatively calculate the ratio of mass be accreted to the black hole to the mass falls back,
\begin{equation}
f_{\rm BH} = \frac{M_{\rm accreted}}{M_{\rm fallback}}=\frac{\int_0^{205 \rm days} \dot M \ dt}{\int_0^{205 \rm days} \dot M_{\rm inject} \ dt} = 0.15
\end{equation}
Only $15\%$ of the fallback debris mass is accreted to the black hole.

\begin{figure}
\begin{center}
\includegraphics[scale=0.4]{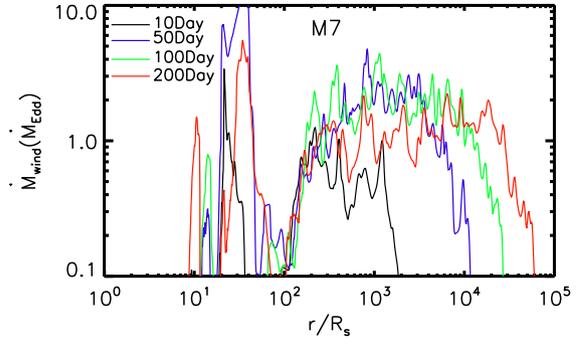}\hspace*{0.1cm}
\hspace*{0.5cm} \caption{Radial profiles of the wind mass flux for model M7. The black, blue, green and red lines correspond to $t= 10$ Day, 50 Day, 100 Day and 200 Day, respectively. \label{fig:moutM7}}
\end{center}
\end{figure}

For radiation pressure dominated accretion flow, winds are common phenomenon (\cite{Curd2019}; \cite{Dai2018}). We also find that strong winds are present in our simulations. In Figure \ref{fig:moutM7}, we show the radial profiles of the wind mass flux at four snapshots. At each snapshot, there is a bump in the wind mass flux in the region $10-100 R_s$. We inject the fallback debris in this region, the calculation of the wind mass flux is quite affected. So the bumps should not be take seriously. We pay attention to the region $r > 100 R_s$. This region is far away from the gas injection region and initially there is no gas at all. All of the wind in the region $r> 100R_s$ is from the region inside 100$R_s$. We can see that the wind head moves outwards with time. At $t=10$ day, the wind head is located roughly at 2000 $R_s$. At $t=50$ day, the wind head has arrived at roughly at $10^4 R_s$. Finally, at $t=200$ day, the wind head arrives at $6 \times 10^4 R_s$. At the final point of our simulation, the wind head has not arrived at the outer boundary of the simulation. In future, it is interesting to study how the winds evolve at a much larger scale.

\begin{figure}
\begin{center}
\includegraphics[scale=0.43]{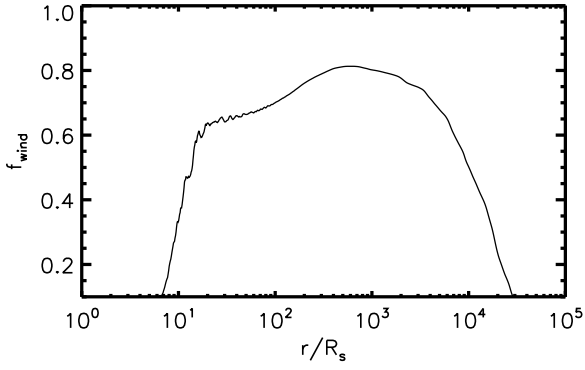}\hspace*{0.1cm}
\hspace*{0.5cm} \caption{Radial profile of the ratio of time integrated mass taken away by wind to the mass of the fallback debris for model M7. \label{fig:mwindM7}}
\end{center}
\end{figure}

\begin{figure}
\begin{center}
\includegraphics[scale=0.4]{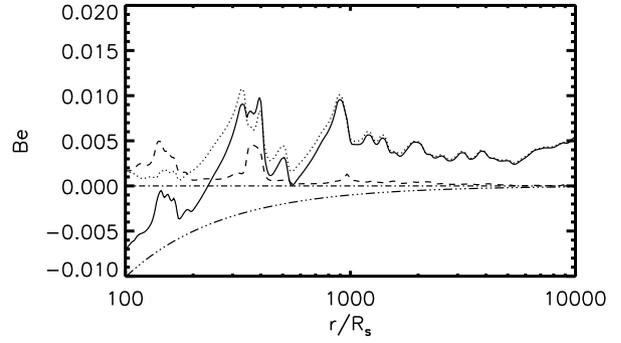}\hspace*{0.1cm}
\hspace*{0.5cm} \caption{Radial profile of the Bernoulli parameter along the midplane at $t = 200$ Day for model M7. The solid line shows the Bernoulli parameter. The dotted, dashed, dotted-dashed and dot-dot-dot-dashed lines correspond to the kinetic energy, gas enthalpy, radiation energy enthalpy and the gravitational potential, respectively. The Bernoulli parameter is calculated in the code unit with $GM_{\rm BH} = R_{\rm s} =1$.  \label{fig:BeM7}}
\end{center}
\end{figure}

It is interesting to ask how much of the fallback debris mass is taken away by wind. We quantitatively calculated the ratio of the time integrated mass taken away by wind to the mass of the debris falls back,
\begin{equation}
f_{\rm wind}(r) = \frac{M_{\rm wind} (r)}{M_{\rm fallback}}=\frac{\int_0^{205 \rm days} \dot M_{\rm wind} (r) \ dt}{\int_0^{205 \rm days} \dot M_{\rm inject} \ dt}
\end{equation}
The result is shown in Figure \ref{fig:mwindM7}. It can be seen that the value of $f_{\rm wind}$ is a function of radius. It roughly increases with radius inside 500$R_s$. In the region $r > 500 R_s$, it decreases with radius. The results can be understood as follows. We inject gas in the region $8R_s < r < 12 R_s$. An accretion flow forms inside $8R_s$. The mass flux of wind from the accretion flow is quite small (see Figure \ref{fig:moutM7}). Outside the injection region, the wind mass flux is large. The reason for the increase of the value of $f_{\rm wind}$ with radius is as follows. We define winds to have positive Bernoulli parameter. There should be such outflows which have negative Bernoulli parameter. Such outflows have not been recorded as wind. With the outwards motion, the Bernoulli parameter of such outflows increases. Finally, the Bernoulli parameter of some portion of these outflows becomes positive. Then the recorded mass flux of winds increases with radius. Therefore, the wind mass flux increases with radius. We take the Bernoulli parameter along the midplane at $t=200$ day as an example to illustrate this point. Figure \ref{fig:BeM7} plots the radial profile of the Bernoulli parameter along the midplane at $t=200$ day. The contribution of the radiation energy to the Bernoulli parameter is zero as shown in this figure. This is because that at this snapshot, the photosphere $\tau_{\rm es} = 1$ at the midplane is located inside 100$R_s$. However, we find that even in the optically thin region $r>100R_s$, the radiation pressure is $\sim 15\%$ of the gravity. The continue acceleration of gas by radiation pressure makes the Bernoulli parameter having a transition from a negative value to a positive value at $230 R_s$. We note that numerical simulations for hot accretion flow \citep{Yuan2015} and super-Eddington accretion flow \citep{Yang2023} all find that with the outward motion of wind, the Bernoulli parameter can increases. The negative Bernoulli parameter of some outward moving gas can become positive at some larger location due to the acceleration of gas. We find that at $\sim 500 R_s$, $f_{\rm wind}$ reaches its maximum value of 0.813. Therefore, more than $80\%$ percents of the fallback debris escapes. This is consistent with the above conclusion that roughly $15\%$ of the fallback debris mass goes to the black hole horizon ($f_{\rm BH} = 0.15$).
In the region $r > \sim 500 R_s$, $f_{\rm wind}$ decreases with radius. The reason is as follows. All of the gas is from the injection region. It takes time for wind to arrive at large radii as shown in Figure \ref{fig:moutM7}. Therefore, at large radius, for a period since the beginning of the simulation, there is no wind at all. The larger the radius is, the longer the period will be. Therefore, the value of $f_{\rm wind}$ at larger radii decreases outwards.

\begin{figure}
\begin{center}
\includegraphics[width=0.42\textwidth]{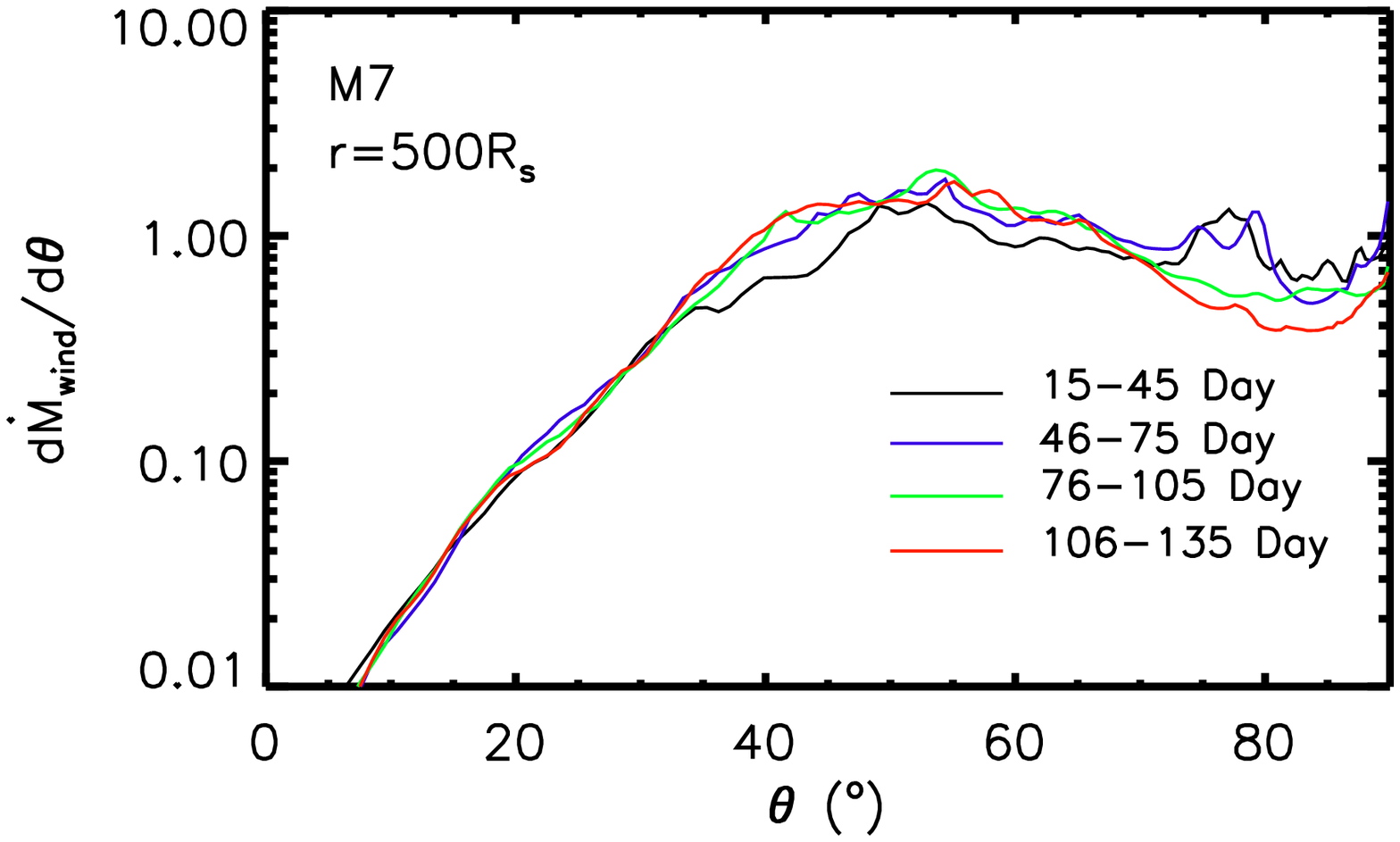}
\includegraphics[width=0.42\textwidth]{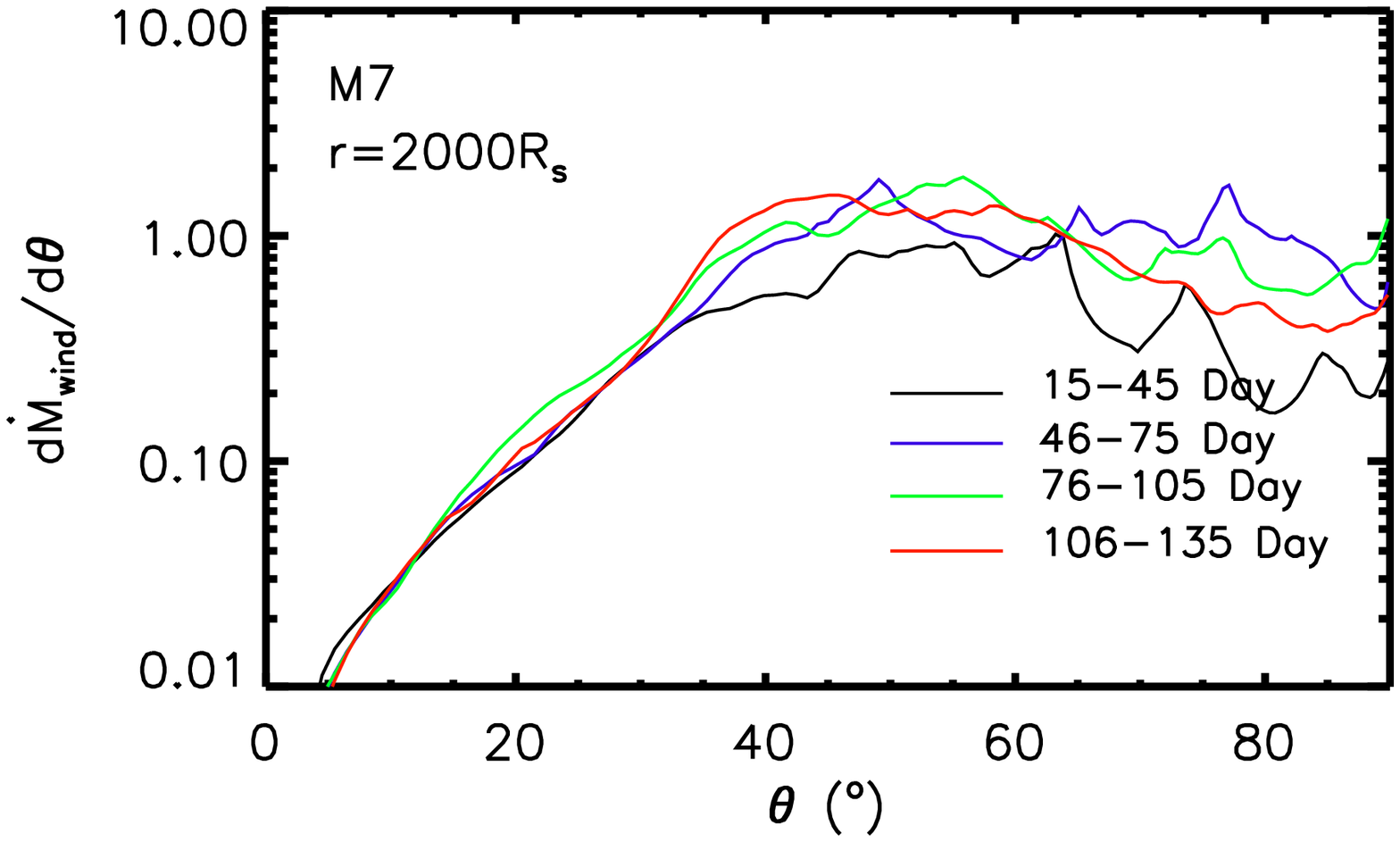}
\caption{The angular ($\theta$) distribution of the mass flux of wind in unit of Eddington accretion rate for model M7. In order to eliminate the fluctuation, we do time-average to the wind mass flux. The black line, blue line, green line and red line correspond to average period of 15-45, 46-75, 76-105 and 106-135 days, respectively. The top panel is for 500$R_s$. The bottom panel is for 2000$R_s$. }
\label{fig:moutthetaM7}
\end{center}
\end{figure}

\begin{figure}
\begin{center}
\includegraphics[width=0.42\textwidth]{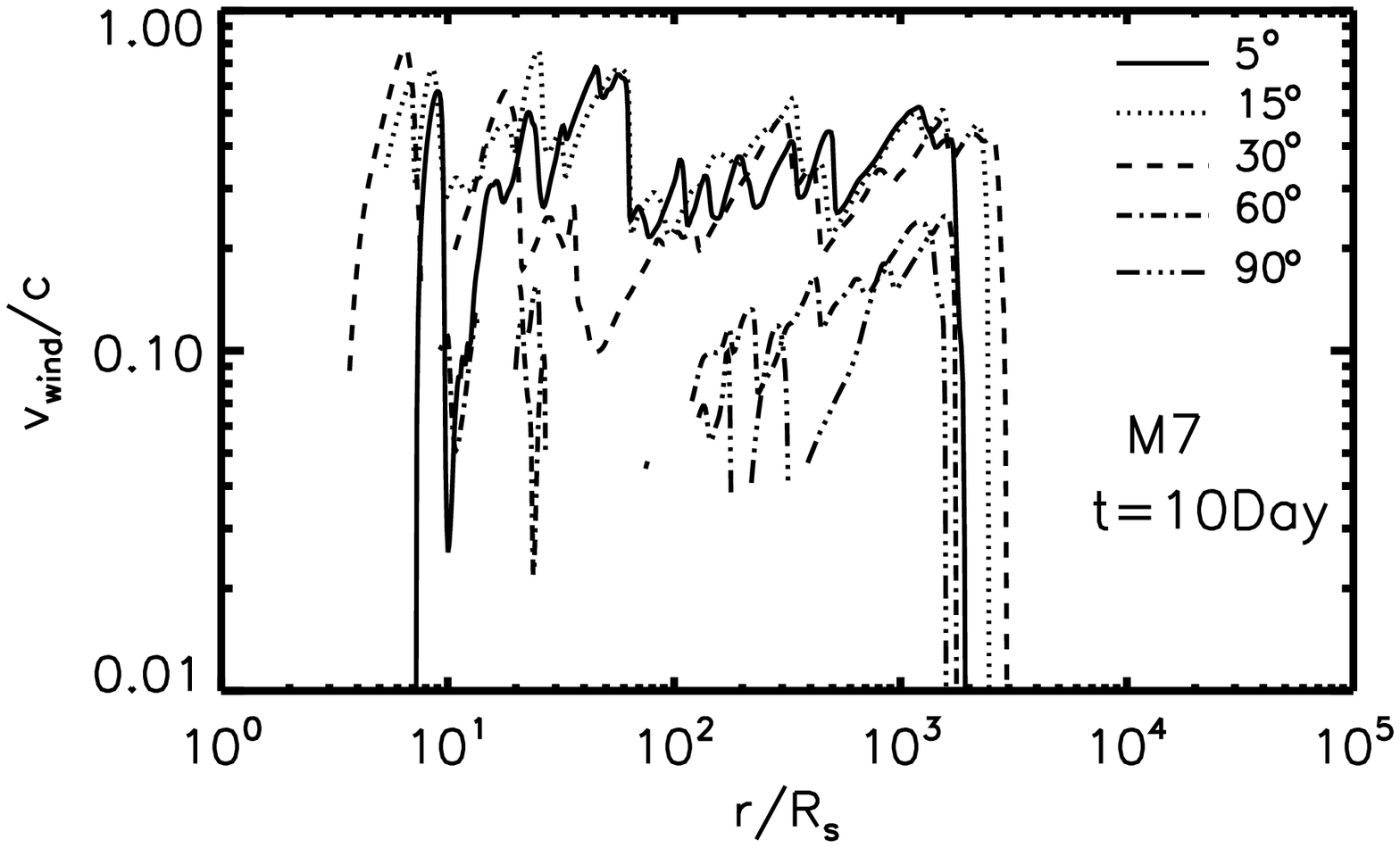}
\includegraphics[width=0.42\textwidth]{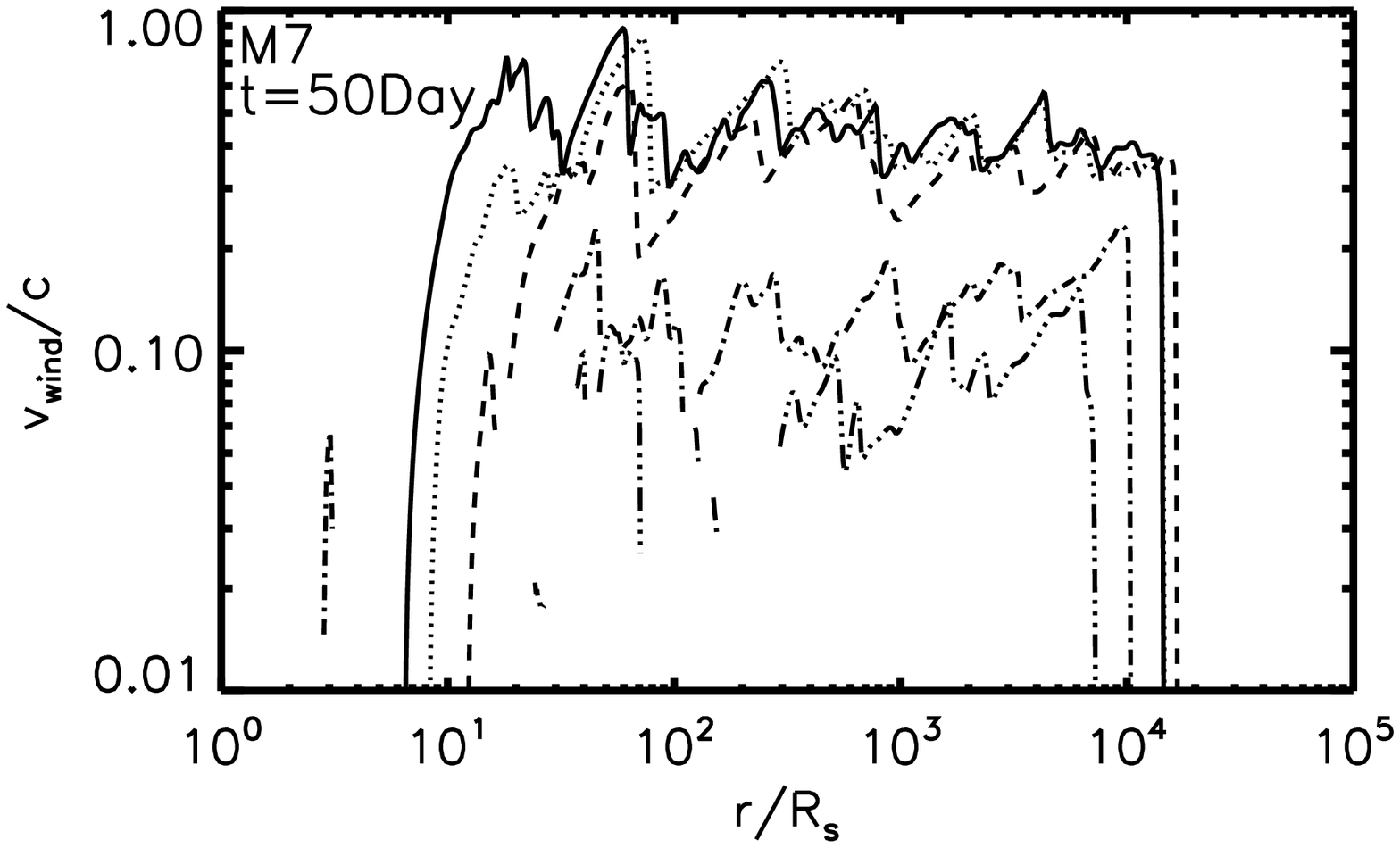} \\
\includegraphics[width=0.42\textwidth]{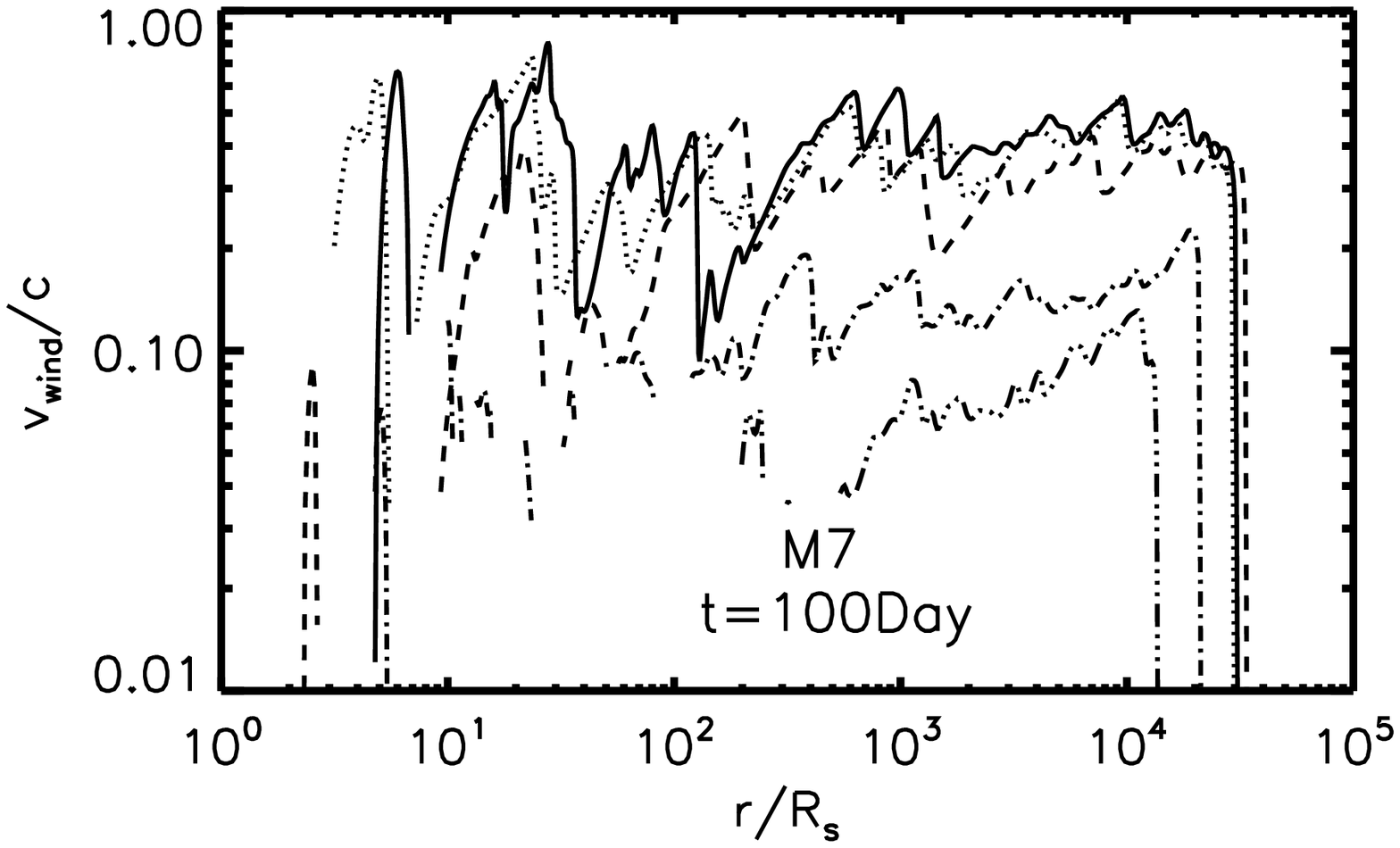}
\includegraphics[width=0.42\textwidth]{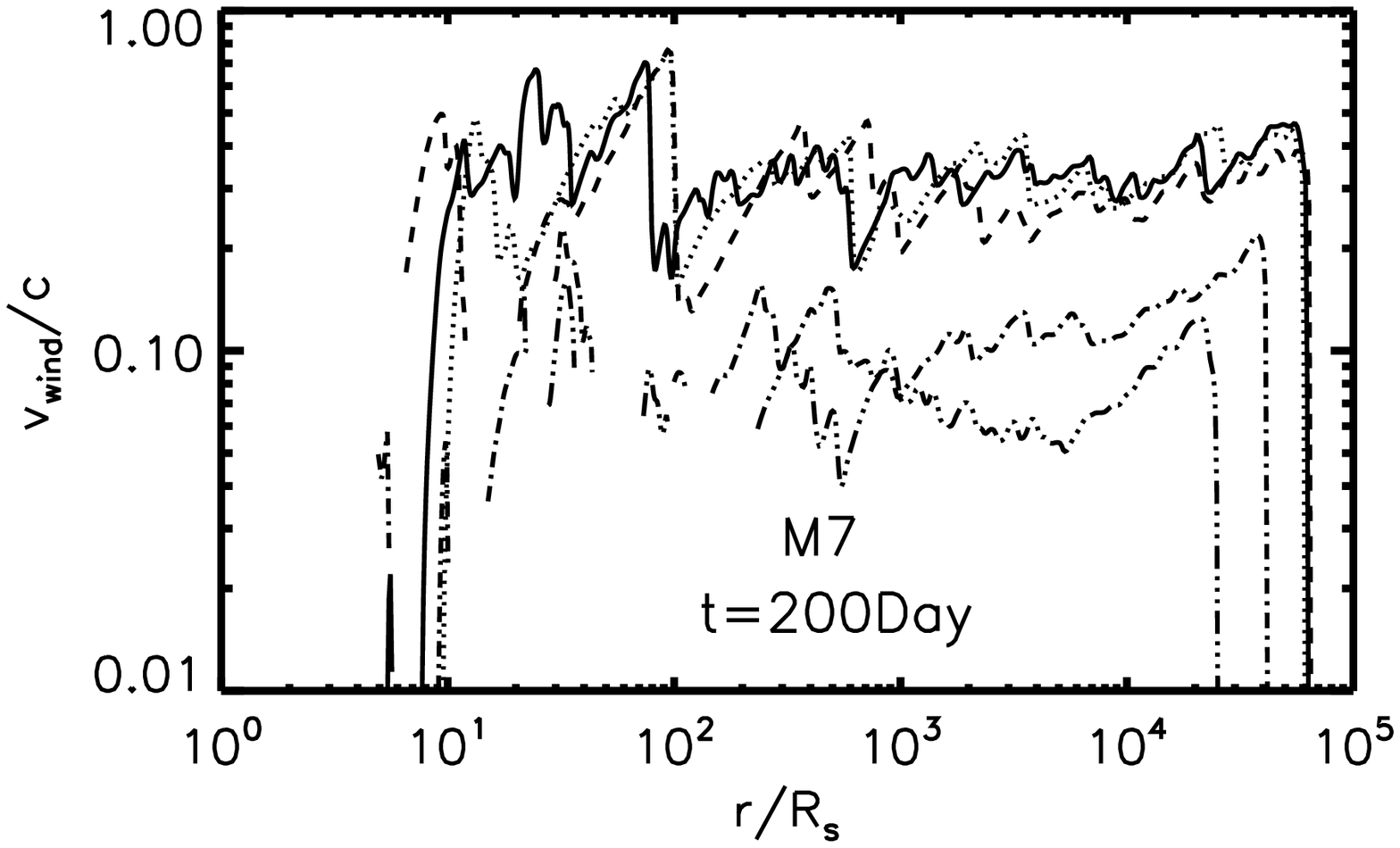}
\caption{The radial profile of the radial velocity of wind for model M7. From top to bottom, the panels correspond to $t= $10 day, 50 day, 100 day, and 200 day, respectively. In each panel, we plot the velocity along 5 viewing angles. }
\label{fig:vrthetaM7}
\end{center}
\end{figure}

The angular ($\theta$) direction distribution of the mass flux of wind is shown in Figure \ref{fig:moutthetaM7}. In order to eliminate the fluctuation, we do time-average to the wind mass flux. The black line, blue line, green line and red line correspond to average period of 15-45, 46-75, 76-105 and 106-135 days, respectively. The top panel is for 500$R_s$. The bottom panel is for 2000$R_s$. It is clear that close to the rotational axis, the mass flux of wind is lowest. The mass flux of wind close to the rotational axis is more than 2 orders of magnitude lower than that in the angular region of $\theta > 40^\circ$. The mass flux of wind increases from $\theta = 0^\circ$ to $\theta=40^\circ$. Recent numerical simulations of super-Eddington accretion flow also found the similar angular distribution of mass flux of winds \citep{Yang2023}. The low gas density close to the rotational axis results in the low mass flux of wind there. In the region $40^\circ < \theta < 90^\circ$, the mass flux of wind is roughly a constant with $\theta$ angle.

TDE winds are probably responsible for the radio emission in some TDEs (see \cite{Alexander2020} for review). The winds can interact with the CNM \citep{Barniol2013} or dense cloud surrounding the black hole (\cite{Mou2022}; \cite{Bu2023}), which induces shock. The power law electrons responsible for radio emission can be accelerated in shock. In the shock model, the very important two parameters are the velocity and the kinetic power of winds. Therefore, it is very important to give the velocity and kinetic power of TDEs winds by simulations.

\begin{figure}
\begin{center}
\includegraphics[scale=0.42]{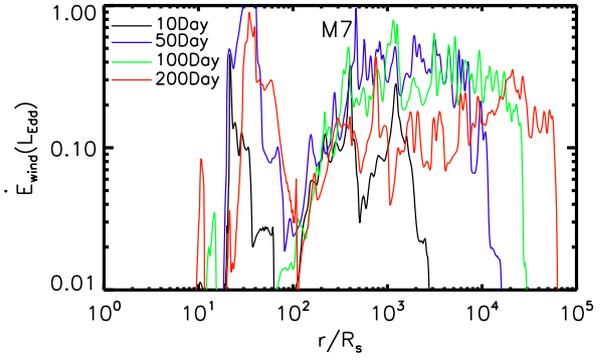}\hspace*{0.1cm}
\hspace*{0.5cm} \caption{Radial profiles of the kinetic power of wind for model M7. The black, blue, green and red lines correspond to $t= 10$ Day, 50 Day, 100 Day and 200 Day, respectively. \label{fig:powerM7}}
\end{center}
\end{figure}

In Figure \ref{fig:vrthetaM7} we plot the radial profile of the radial velocity of wind along several viewing angles. It is clear that generally, the velocity of wind decreases from the rotational axis towards the midplane. Close to the rotational axis, the maximum velocity of wind can achieve $0.7 c$. At the midplane, the velocity of wind is roughly $0.1 c$. The decrease of wind velocity from the rotational axis towards the midplane is a common phenomenon in both radiation pressure dominated super-Eddington accretion flow \cite{Yang2023} and low accretion rate hot accretion flow \cite{Yuan2015}. At $t=10 $ day, the wind along the viewing angles of $ \theta < 30^\circ$ arrives at $\sim 2000R_s$, which is the distance the wind moving with a velocity of $\sim 0.5 c$ in 10 days. Along a fixed viewing angle ( especially in the region $\theta < 30^\circ$ ), the velocity of winds is roughly a constant with radius. This means that the velocity of wind is much larger than the escape velocity, the gravity can hardly decelerate the wind. With the roughly constant velocity, at the end of the simulation $200$ day, the wind arrives at $r \sim 6 \times 10^4 R_s \sim 1.8 \times 10^{17} $cm. We note that in our simulations, we do not consider the deceleration of wind by the CNM or dense cloud. In future, it is interesting to simulate a more realistic case in which the CNM or dense cloud is properly considered.

\begin{figure}
\begin{center}
\includegraphics[width=0.42\textwidth]{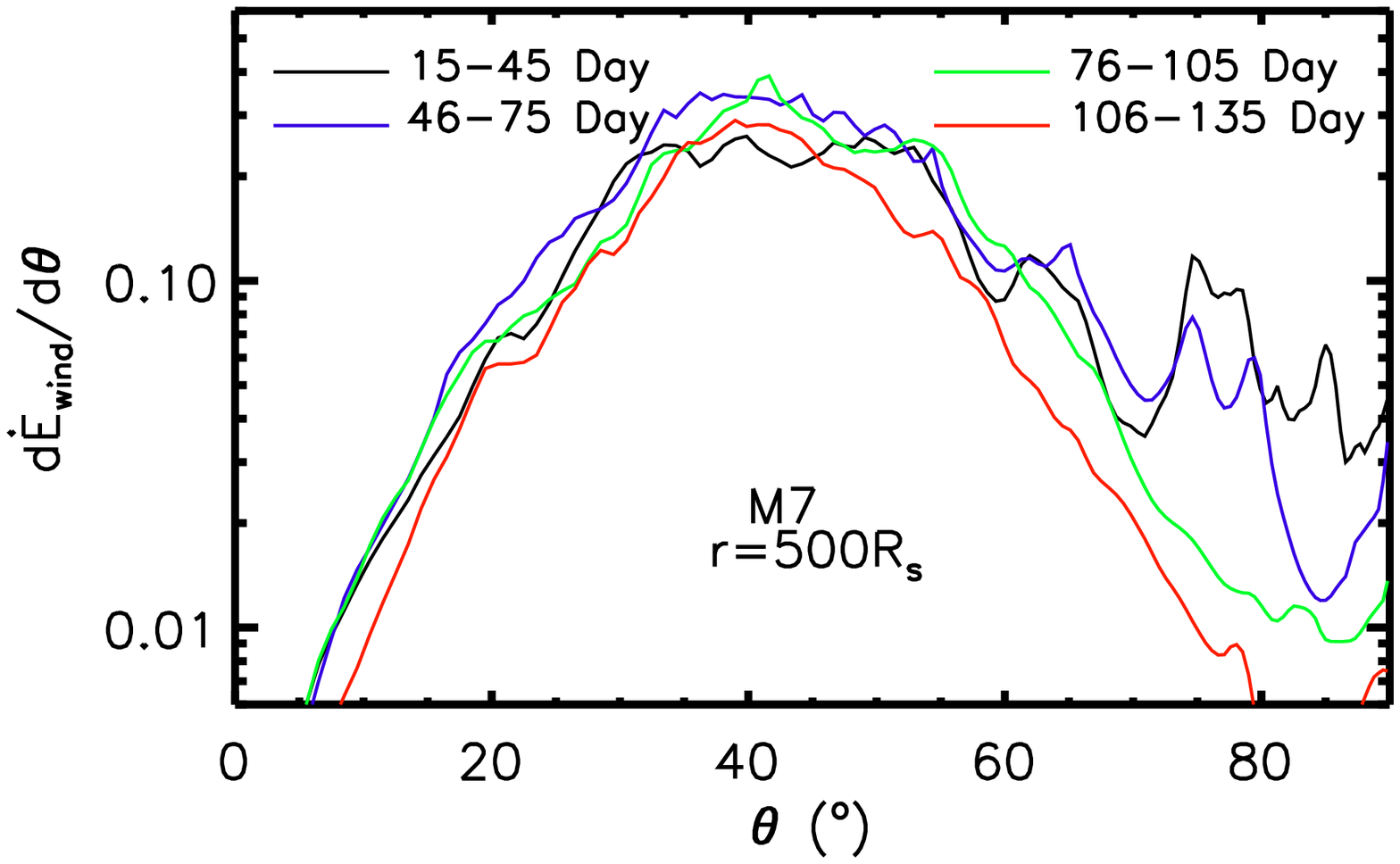}
\includegraphics[width=0.42\textwidth]{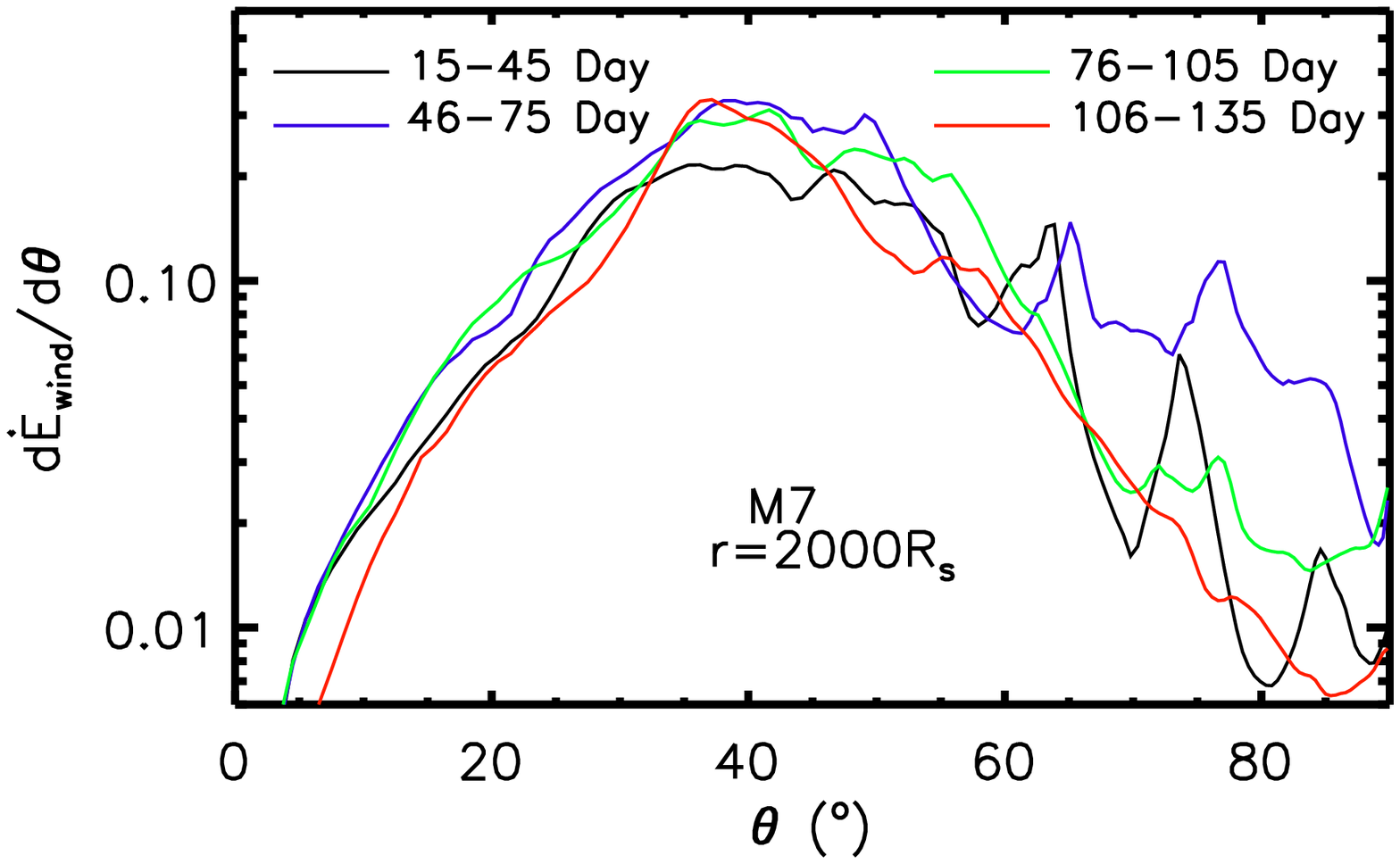}
\caption{The angular ($\theta$) distribution of the kinetic power of wind in unit of $L_{\rm Edd}$ for model M7. In order to eliminate the fluctuation, we do time-average to the kinetic power. The black line, blue line, green line and red line correspond to average period of 15-45, 46-75, 76-105 and 106-135 days, respectively. The top panel is for 500$R_s$. The bottom panel is for 2000$R_s$. }
\label{fig:powerthetaM7}
\end{center}
\end{figure}

In figure \ref{fig:powerM7}, we plot the radial profile of the kinetic power of wind at four snapshots. There are bumps in the region $10 R_s < r < 100 R_s$.  We also find bumps in the radial profile of mass flux of wind above (see Figure \ref{fig:moutM7}). As introduced above, the bumps are related to the wind injection in this region. We pay attention to wind at much larger radii $r > 100 R_s$, where is hardly affected by the wind injection. At $t = 10 $ day, the wind moves to roughly $ r \sim 2000R_s$ (see the top left panel of Figure \ref{fig:vrthetaM7}), therefore, we can see that for this snapshot, the kinetic power of wind outside 2000 $R_s$ is cutoff. With the increase of time, the cutoff radius of the kinetic power of wind increases due to the outwards movement of wind. The kinetic power of wind can be $ > 0.5 L_{\rm Edd} \sim 6.5 \times 10^{44} {\rm erg \ s^{-1}}$. The kinetic power of wind is enough to account for most of the radio emissions in radio TDEs (\cite{Mou2022}; \cite{Bu2023}).

The angular distribution (or opening angle) of wind is an important parameter for the study of interaction between wind and CNM. In Figure \ref{fig:powerthetaM7}, we show the angular distribution of the kinetic power of wind. We do time-average to eliminate the fluctuation. The black line, blue line, green line and red line correspond to average period of 15-45, 46-75, 76-105 and 106-135 days, respectively. The top panel is for 500$R_s$. The bottom panel is for 2000$R_s$. It can been seen that at both radii, the kinetic power is largest in the region of $30^\circ < \theta < 50^\circ$. Also, in this region, the kinetic power of wind is almost a constant with $\theta$. In the region $\theta < 30^\circ$, with the decrease of $\theta$, the kinetic power decreases very quickly. The velocity of wind in this region is highest (see Figure \ref{fig:vrthetaM7}). However, the mass flux of wind is lowest (see Figure \ref{fig:moutthetaM7}). The low mass flux in this region results in the low kinetic power. In the region of $\theta > 50^\circ$, the kinetic power of wind decreases quickly with increase of $\theta$ due to both the quick decrease of velocity with $\theta$ (see Figure \ref{fig:vrthetaM7}) and the slow decrease of mass flux with $\theta$ (see Figure \ref{fig:moutthetaM7}).

\subsection{Model M6}

\begin{figure}
\begin{center}
\includegraphics[width=0.42\textwidth]{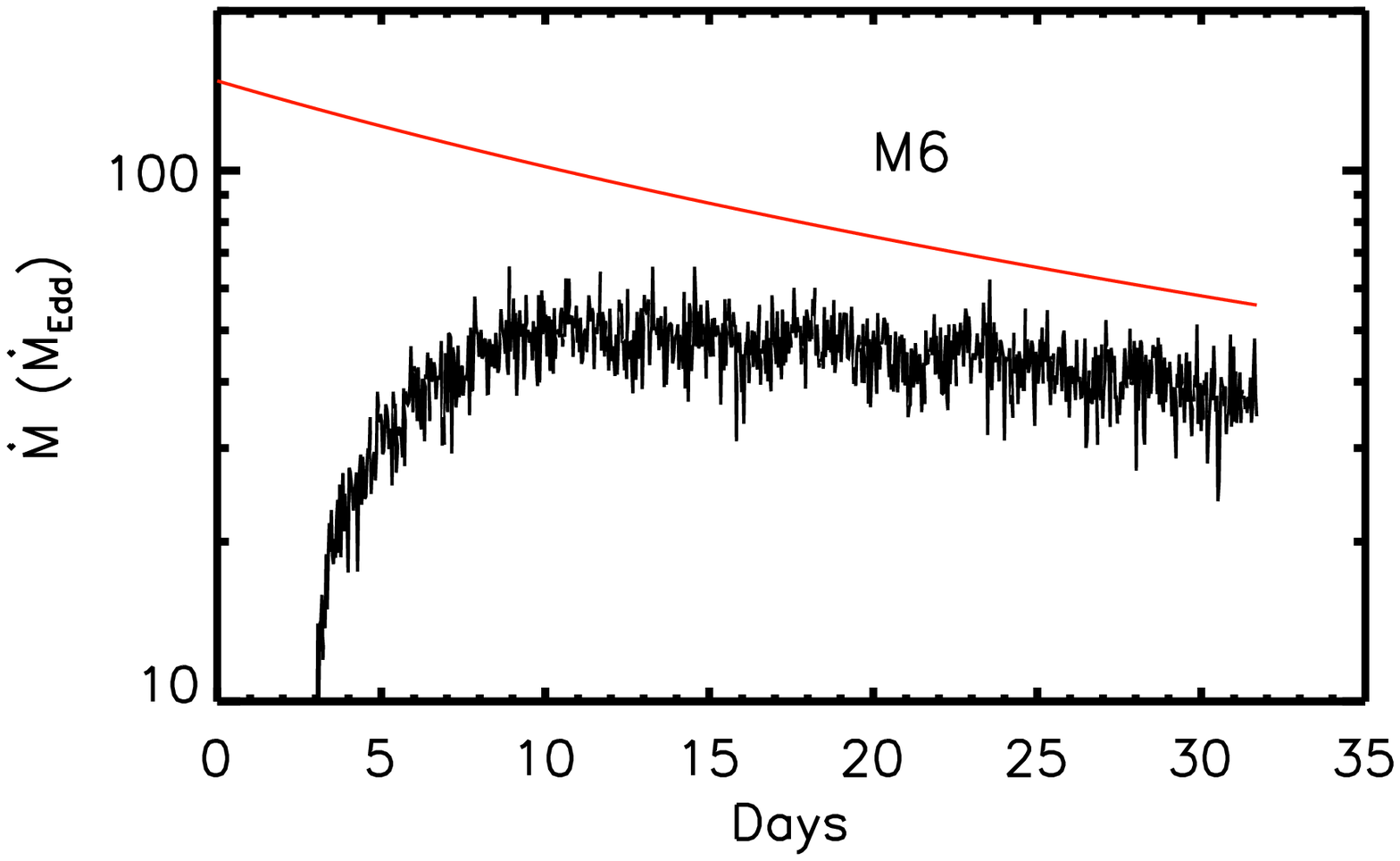}
\includegraphics[width=0.42\textwidth]{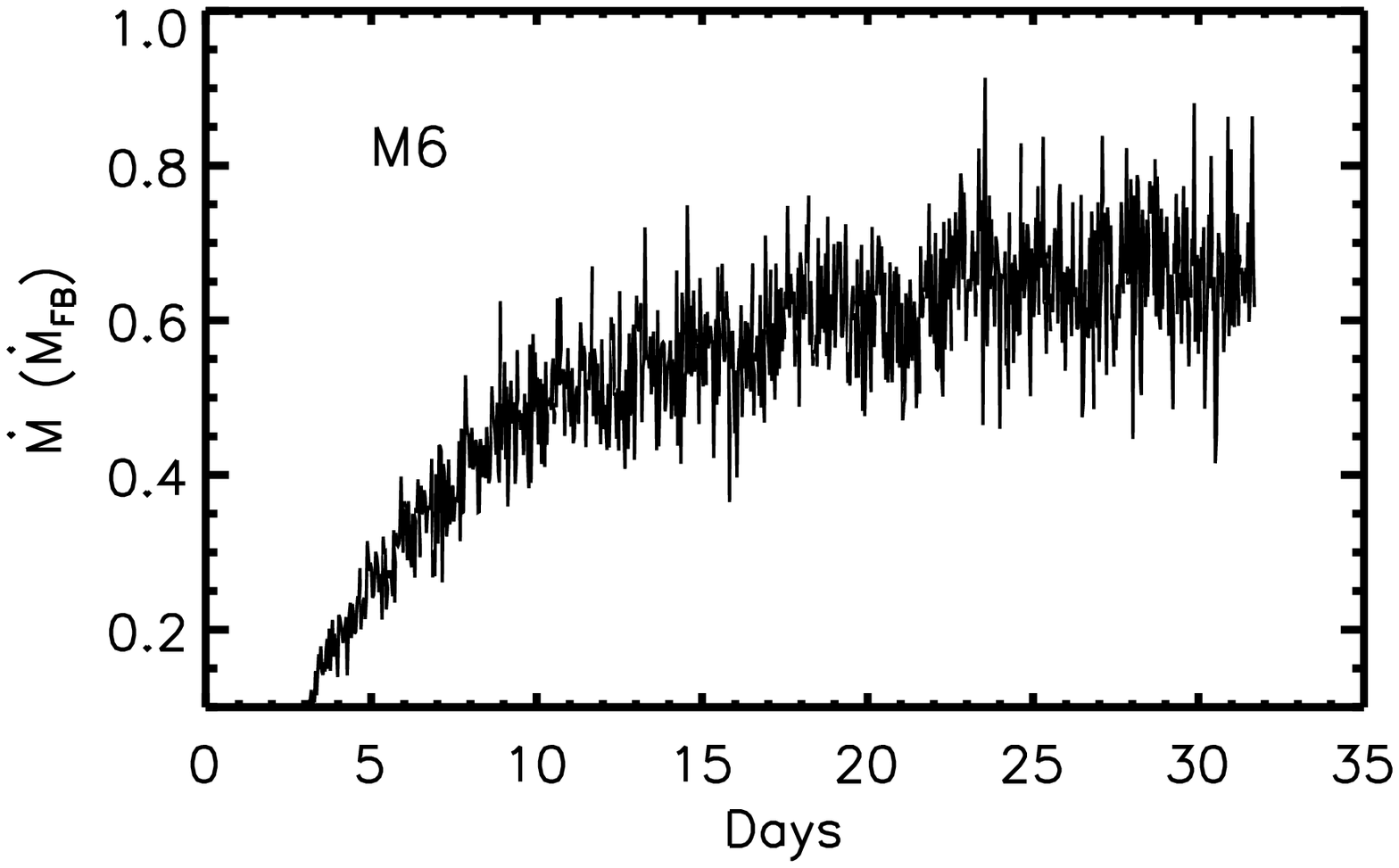}
\caption{Accretion rate for model M6. Top panel: time evolution of the black hole accretion rate (black line) and the stellar debris fallback rate (red line) in unit of Eddington accretion rate. Bottom panel: time evolution of the black hole accretion rate in unit of the stellar debris fallback rate.}
\label{fig:mdotM6}
\end{center}
\end{figure}

\begin{figure}
\begin{center}
\includegraphics[scale=0.43]{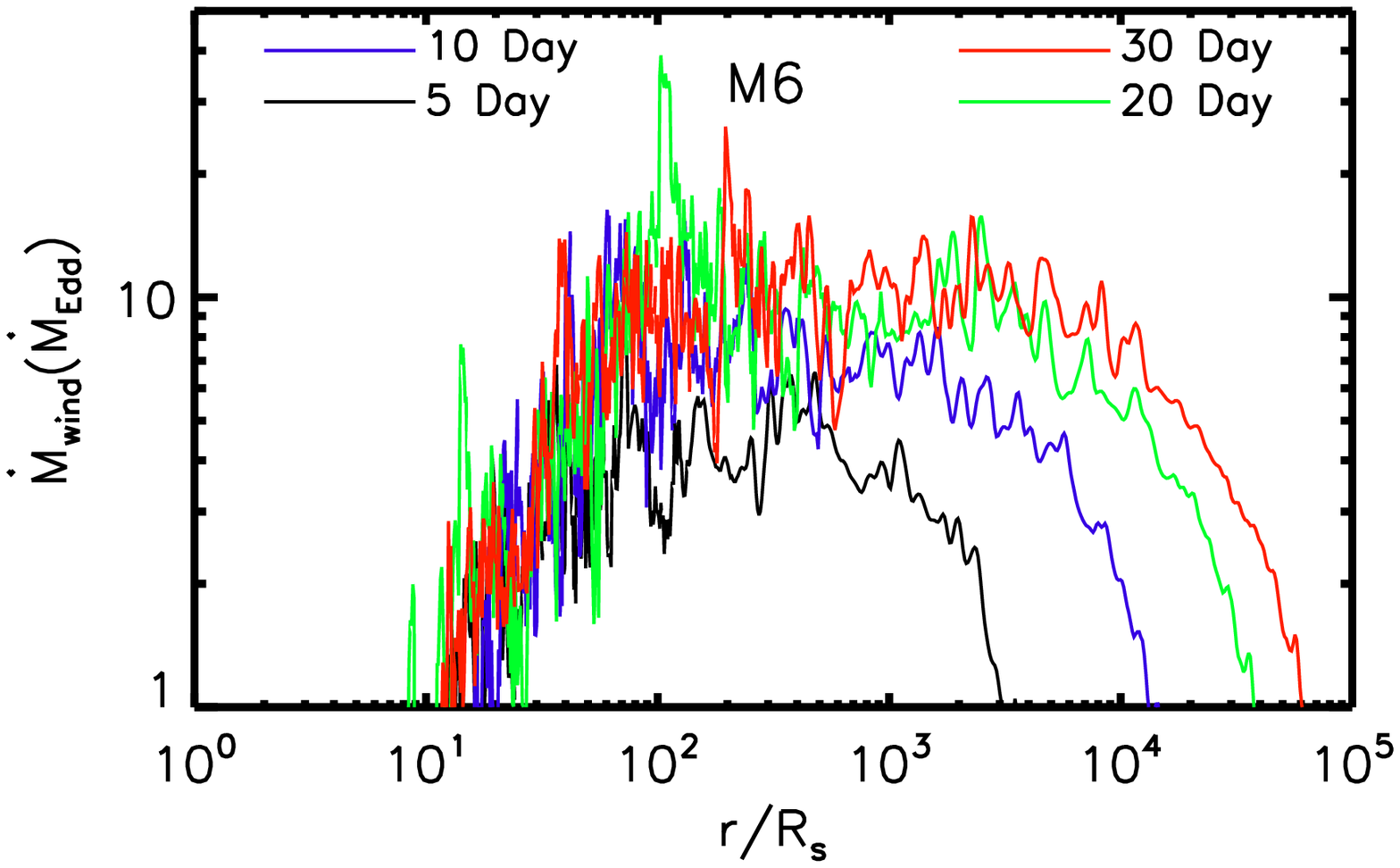}\hspace*{0.1cm}
\hspace*{0.5cm} \caption{Radial profiles of the wind mass flux for model M6. The black, blue, green and red lines correspond to $t= 5$ Day, 10 Day, 20 Day and 30 Day, respectively. \label{fig:moutM6}}
\end{center}
\end{figure}

\begin{figure}
\begin{center}
\includegraphics[scale=0.43]{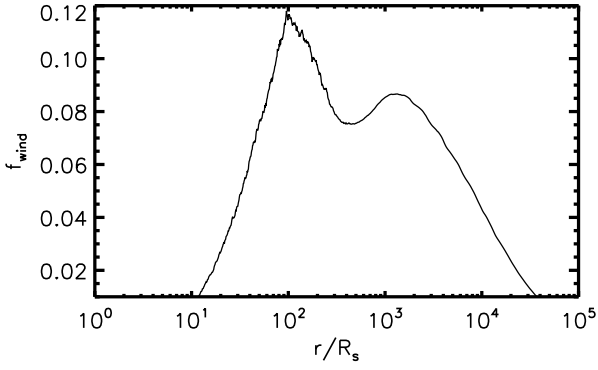}\hspace*{0.1cm}
\hspace*{0.5cm} \caption{Radial profile of the ratio of the time integrated mass taken away by wind to the mass of the fallback debris for model M6. \label{fig:mwindM6}}
\end{center}
\end{figure}

In model M6, the central black hole mass is $10^6M_\odot$. We inject the fallback stellar debris around the circularization radius $47R_s$. The simulation covers 32 days since the peak fallback rate. The reason for the much shorter simulated period compared to model M7 is as follows. The time step ($\Delta t$) of integration of the simulation is determined by the conditions in the innermost region of the grids. The value $\Delta t \sim \Delta r_{\rm min} / c$, with $\Delta r_{\rm min}$ being the smallest grid at the inner radial boundary. The value of $\Delta r_{\rm min}$ in model M6 is 10 times smaller than that in model M7. Therefore, $\Delta t$ in model M6 is 10 times smaller. The CPU time needed for simulating 32 days for model M6 is longer than that for simulating $200$ days for model M7. The fallback rate is super-Eddington. An viscous radiation pressure dominated accretion flow develops and strong winds are found.

\begin{figure}
\begin{center}
\includegraphics[width=0.42\textwidth]{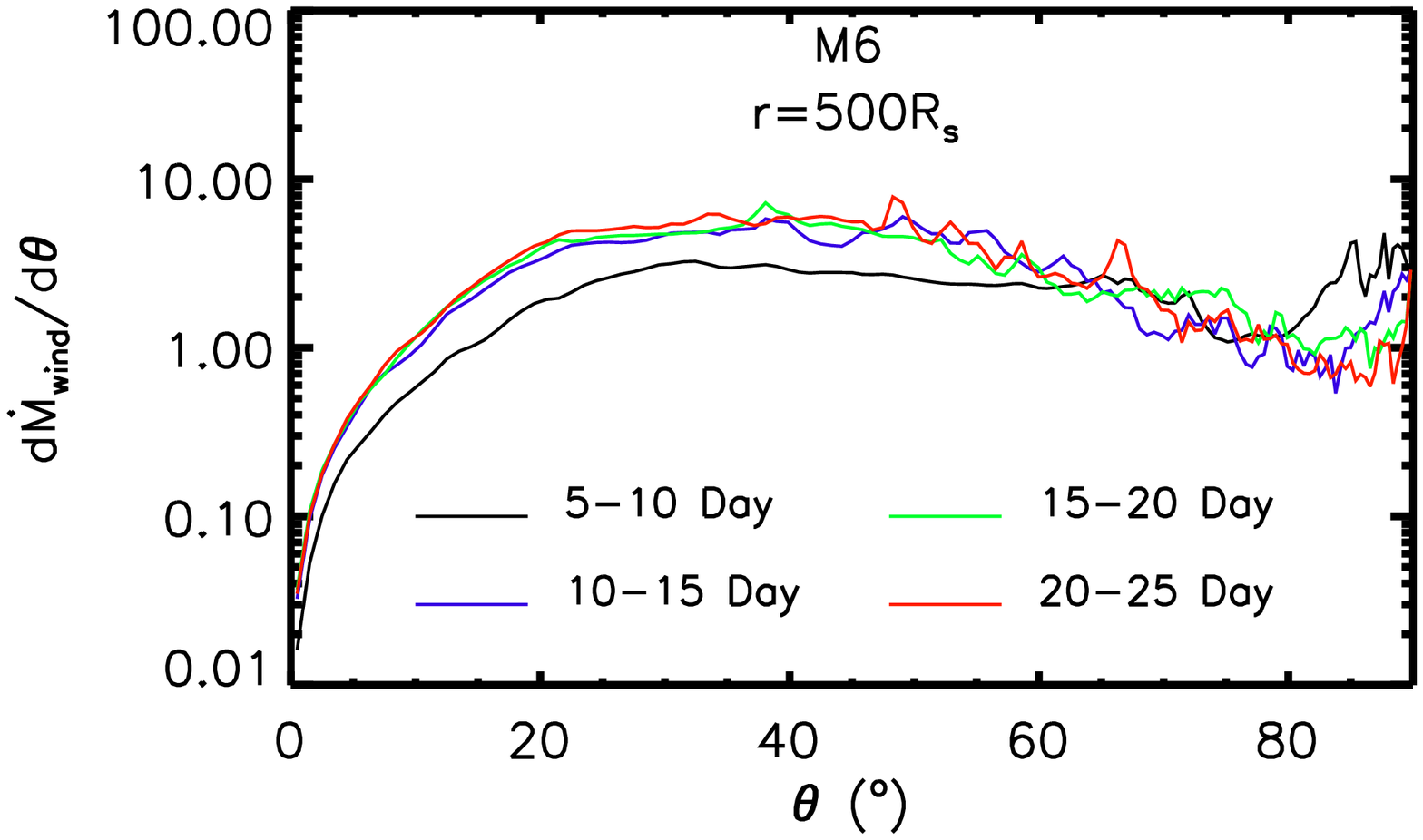}
\includegraphics[width=0.42\textwidth]{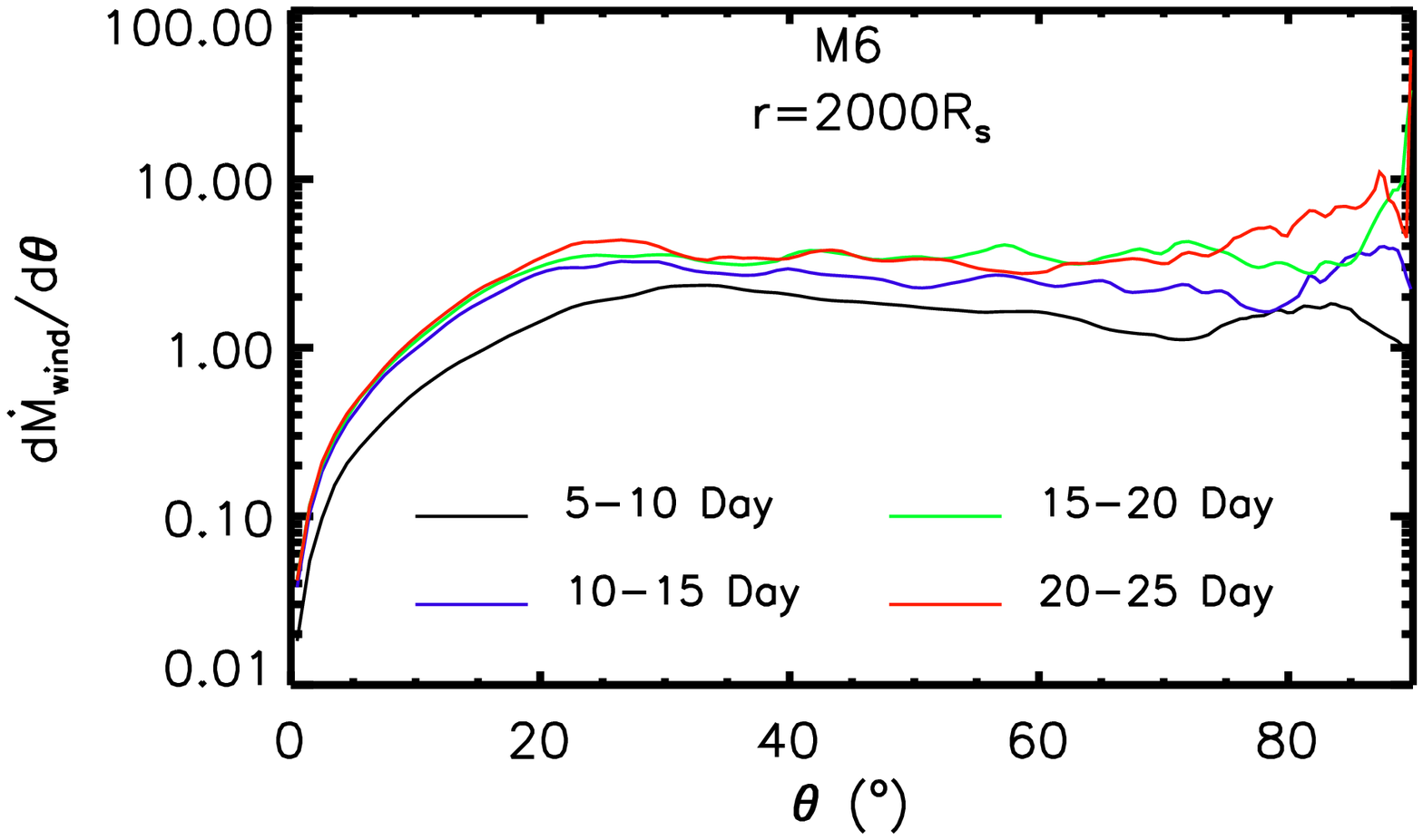}
\caption{The angular ($\theta$) distribution of the mass flux of wind in unit of Eddington accretion rate for model M6. In order to eliminate the fluctuation, we do time-average to the wind mass flux. The black line, blue line, green line and red line correspond to average period of 5-10, 10-15, 15-20 and 20-25 days, respectively. The top panel is for 500$R_s$. The bottom panel is for 2000$R_s$. }
\label{fig:moutthetaM6}
\end{center}
\end{figure}

We first study the mass accretion rate onto the black hole. The top panel of Figure \ref{fig:mdotM6} shows the time evolution of the black hole accretion rate (black line) and the stellar debris fallback rate (red line) in unit of Eddington accretion rate. The ratio of black hole accretion rate to the debris fall back rate is shown in the bottom panel of Figure \ref{fig:mdotM6}. As in model M7, the accretion rate fluctuates with time due to the turbulent motions induced by convective instability. The ratio of accretion rate to the debris fallback rate is higher in model M6 compared to that in model M7 (see Figure \ref{fig:mdotM7}). We also quantitatively calculate the ratio of mass be accreted to the black hole to the mass falls back by using Equation (1). The time-integration is from $t = 0$ to 32 days. We find that in this model,
\begin{equation}
f_{\rm  BH} = 0.43
\end{equation}
$43\%$ of the fallback debris mass is accreted to the black hole. As a note that in model M7, $f_{\rm BH} = 0.15$. In super-Eddington accretion flow, the high scattering optical depth results in that the photons co-moving with the gas in the optically thick region. \cite{Ohsuga2003} found that for super-Eddington accretion flow, the higher the accretion rate, the easier the photons can be trapped. In other words, with the increase of accretion rate, the photons are more easier to be advected into the black hole horizon rather than be advected to lager radii by winds. The wind is relatively weaker in higher accretion rate flow. In model M6, the accretion rate is much higher than that in model M7. Therefore, the wind (in the sense of ratio of wind mass flux to accretion rate) is relatively weaker in model M6.

\begin{figure}
\begin{center}
\includegraphics[width=0.42\textwidth]{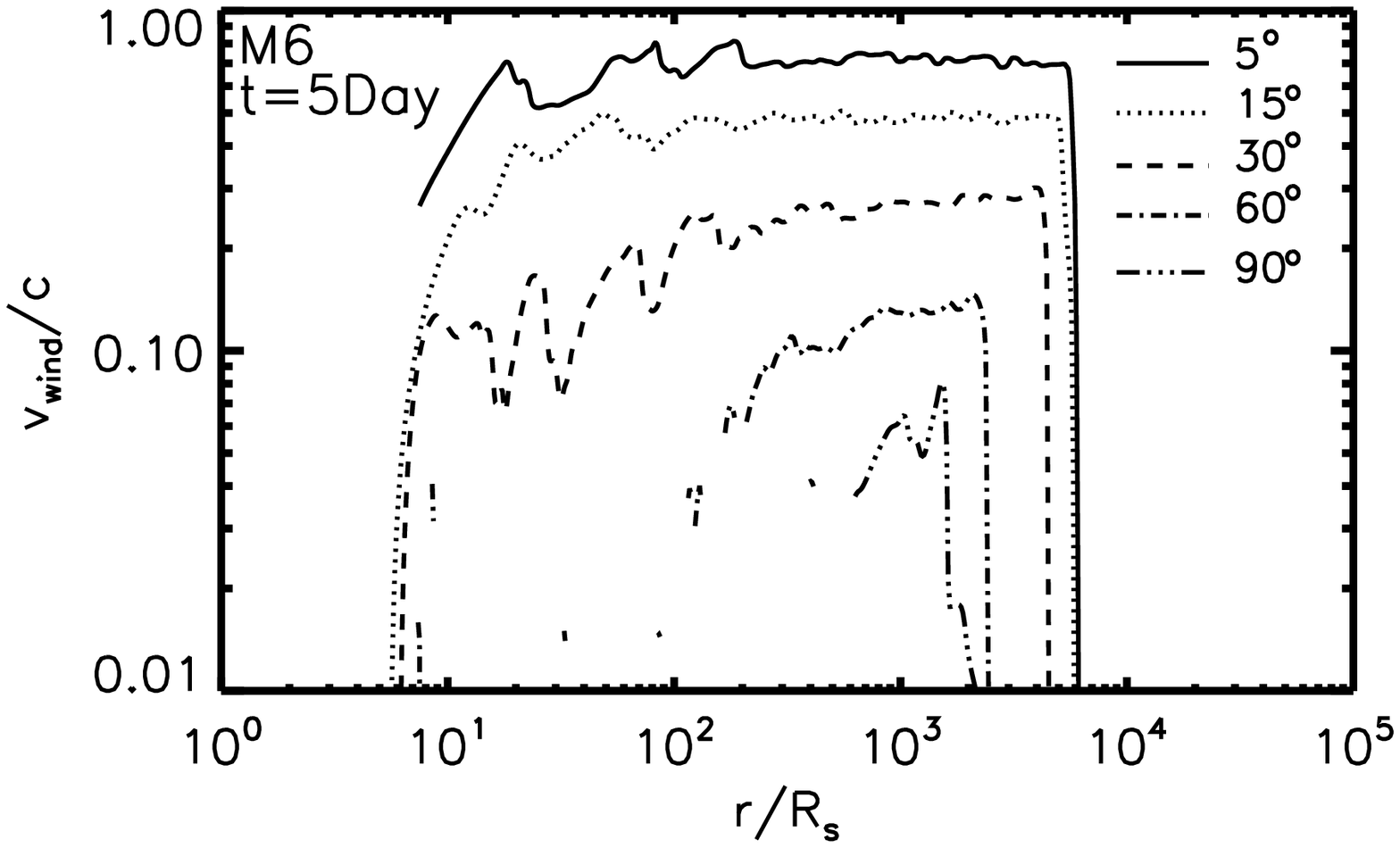}
\includegraphics[width=0.42\textwidth]{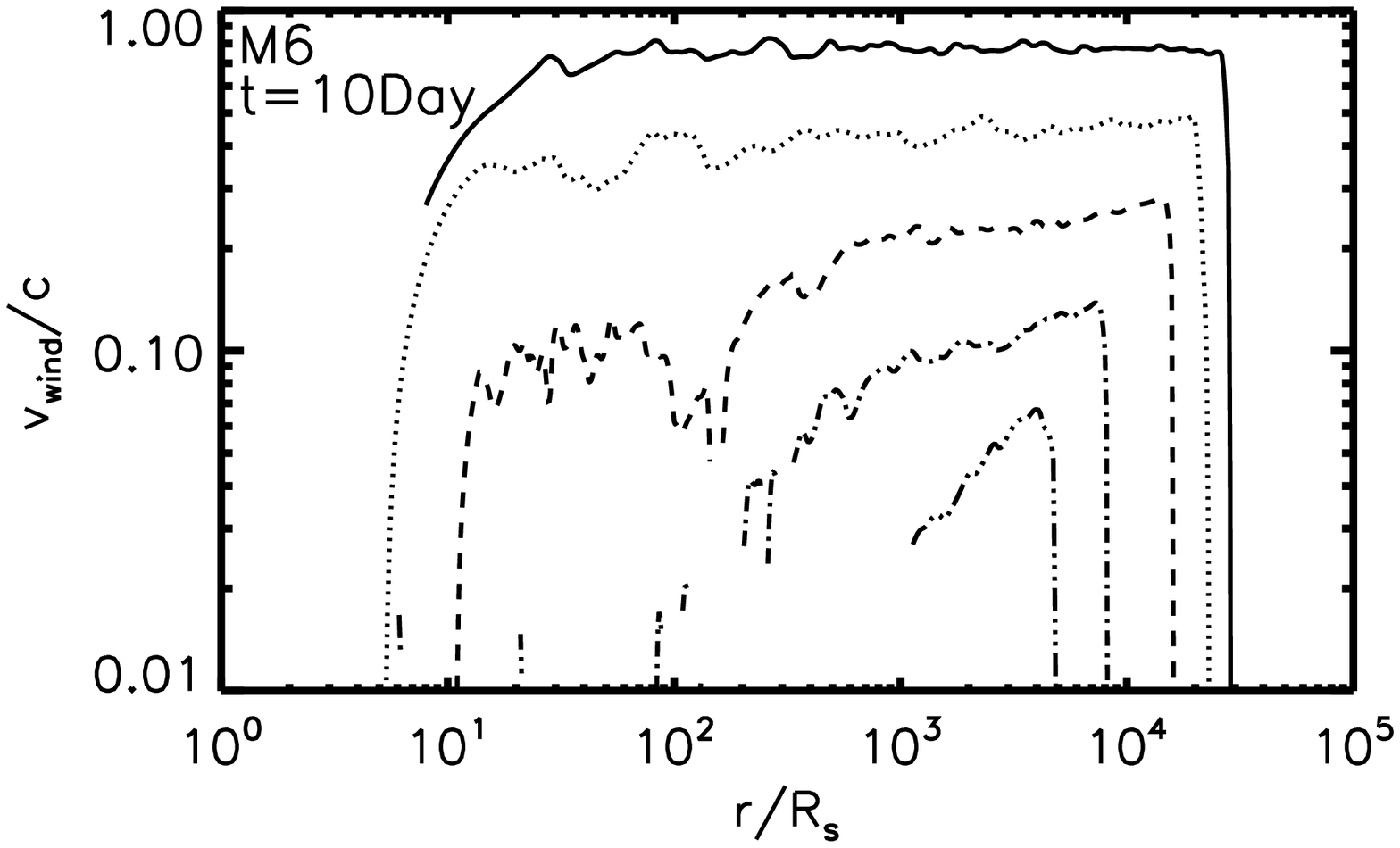} \\
\includegraphics[width=0.42\textwidth]{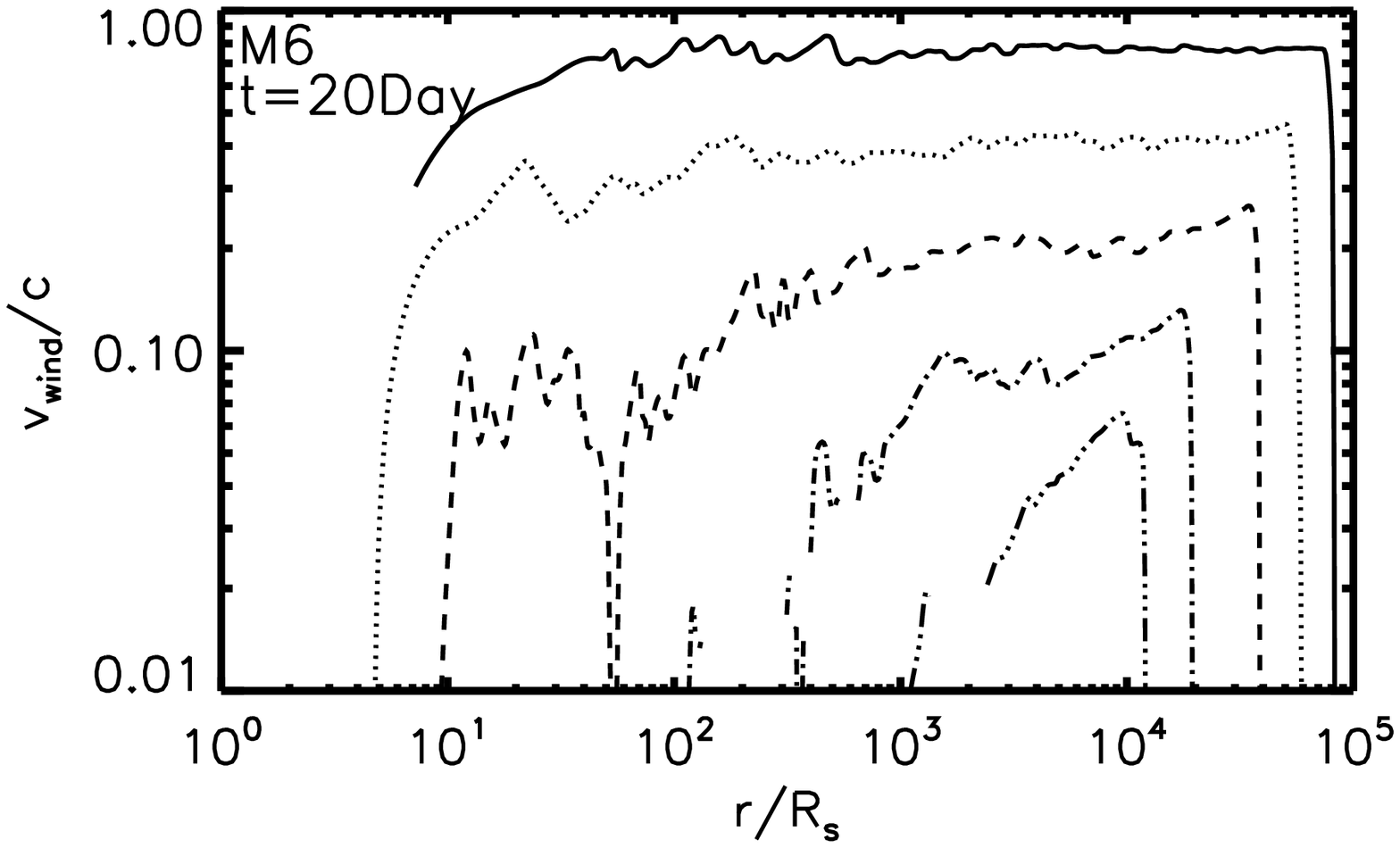}
\includegraphics[width=0.42\textwidth]{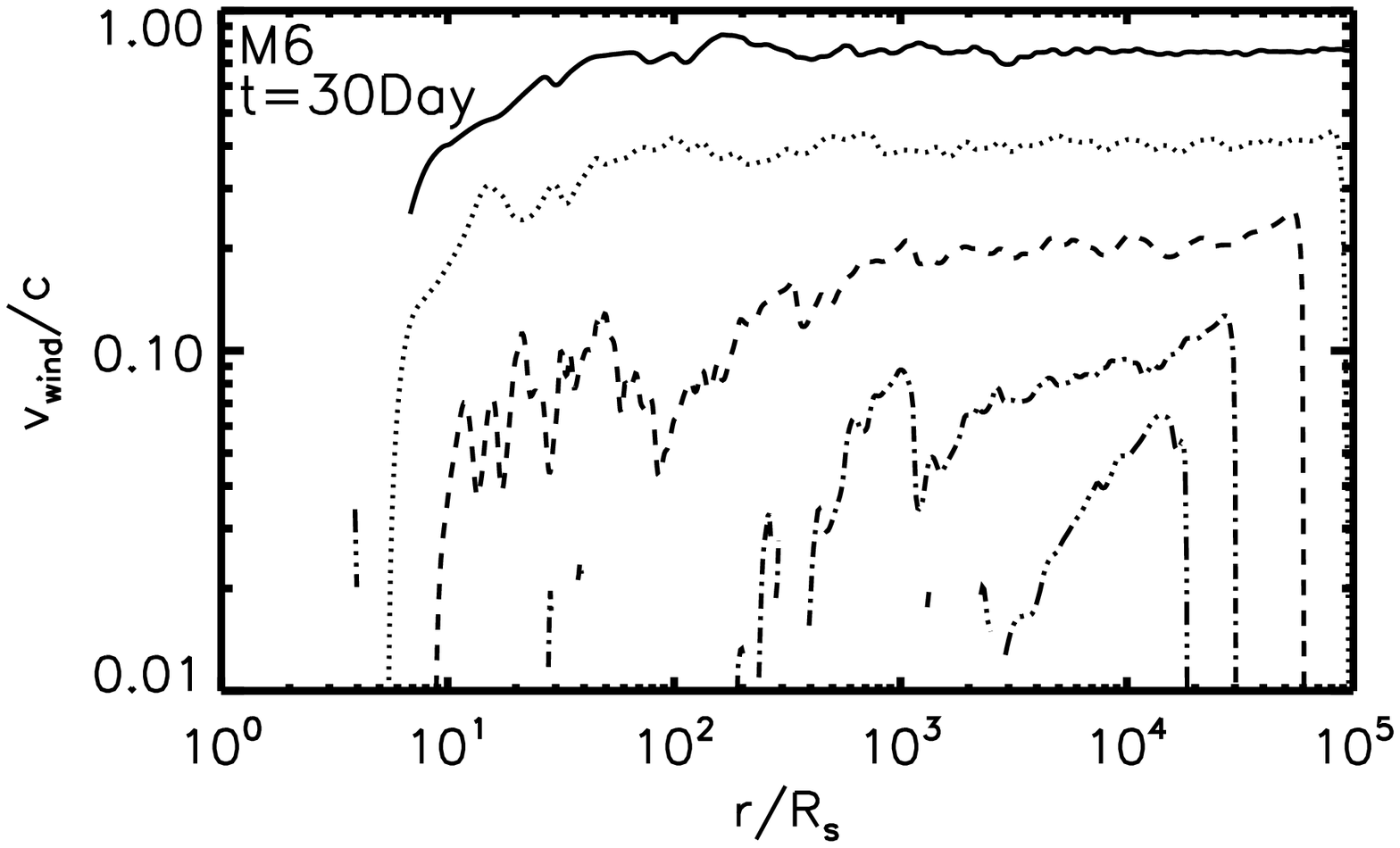}
\caption{The radial profile of the radial velocity of wind for model M6. From top to bottom, the panels correspond to $t= $5 day, 10 day, 20 day, and 30 day, respectively. In each panel, we plot the velocity along 5 viewing angles. }
\label{fig:vrthetaM6}
\end{center}
\end{figure}

We show the radial profiles of the mass flux of wind at four snapshots in Figure \ref{fig:moutM6}. In this model, the injection radii is $47R_s$. An viscous accretion flow forms inside this radius and wind is generated. Inside this radius, we can see that the mass flux of wind increases with radius. This can be understood as follows. In an accretion flow, except the region very close to the black hole, the wind can be generated at any radii. The mass flux of wind at a given radius includes both the flux of wind from the smaller radii and that generated locally. Therefore, we can find that the the wind mass flux increases with radius inside $\sim 47 R_s$. However, outside this radius, all the wind comes from the smaller radii and winds can not be generated locally. Therefore, the mass flux of wind is roughly a constant with radius as shown in Figure \ref{fig:moutM6}. We also see the cutoff of the mass flux of wind as in model M7. The cutoff radius increases with time. At the end of the simulation, the wind arrives at $6\times 10^4R_s$.

\begin{figure}
\begin{center}
\includegraphics[scale=0.43]{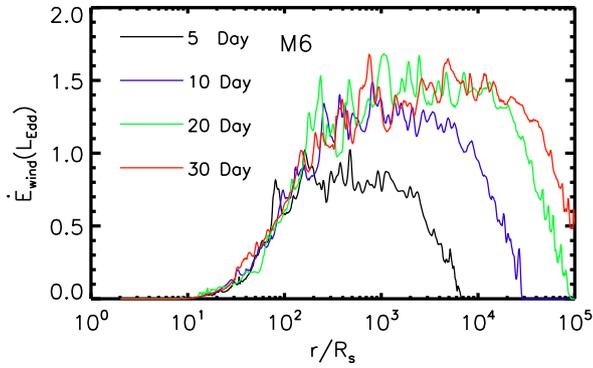}\hspace*{0.1cm}
\hspace*{0.5cm} \caption{Radial profiles of the kinetic power of wind for model M6. The black, blue, green and red lines correspond to $t= 5$ Day, 10 Day, 20 Day and 30 Day, respectively. \label{fig:powerM6}}
\end{center}
\end{figure}

We calculate the mass taken away by wind using Equation (6). The result is shown in Figure \ref{fig:mwindM6}. It is clear that from $10R_s$ to $100R_s$, the mass flux of wind increases quickly. The reason is same as the case for model M7. The wind is defined as outflow with positive Bernoulli parameter. There is outflow with negative Bernoulli parameter. With the outwards motion, the negative Bernoulli parameter of some portion of such outflows becomes positive. Thus, the mass flux of wind increases with radius. Outside 100$R_s$, the mass flux of wind decreases with radius. As explained for Model M7, the wind needs to spend time to arrive at large radius. The larger the radii, the longer the period that there is no wind. Thus, the value of $f_{\rm wind}$ at larger radii decreases outwards. Note that there is a small bump around $1500R_s$. The small bump is due to the variation of Bernoulli parameter of wind.

\begin{figure}
\begin{center}
\includegraphics[width=0.42\textwidth]{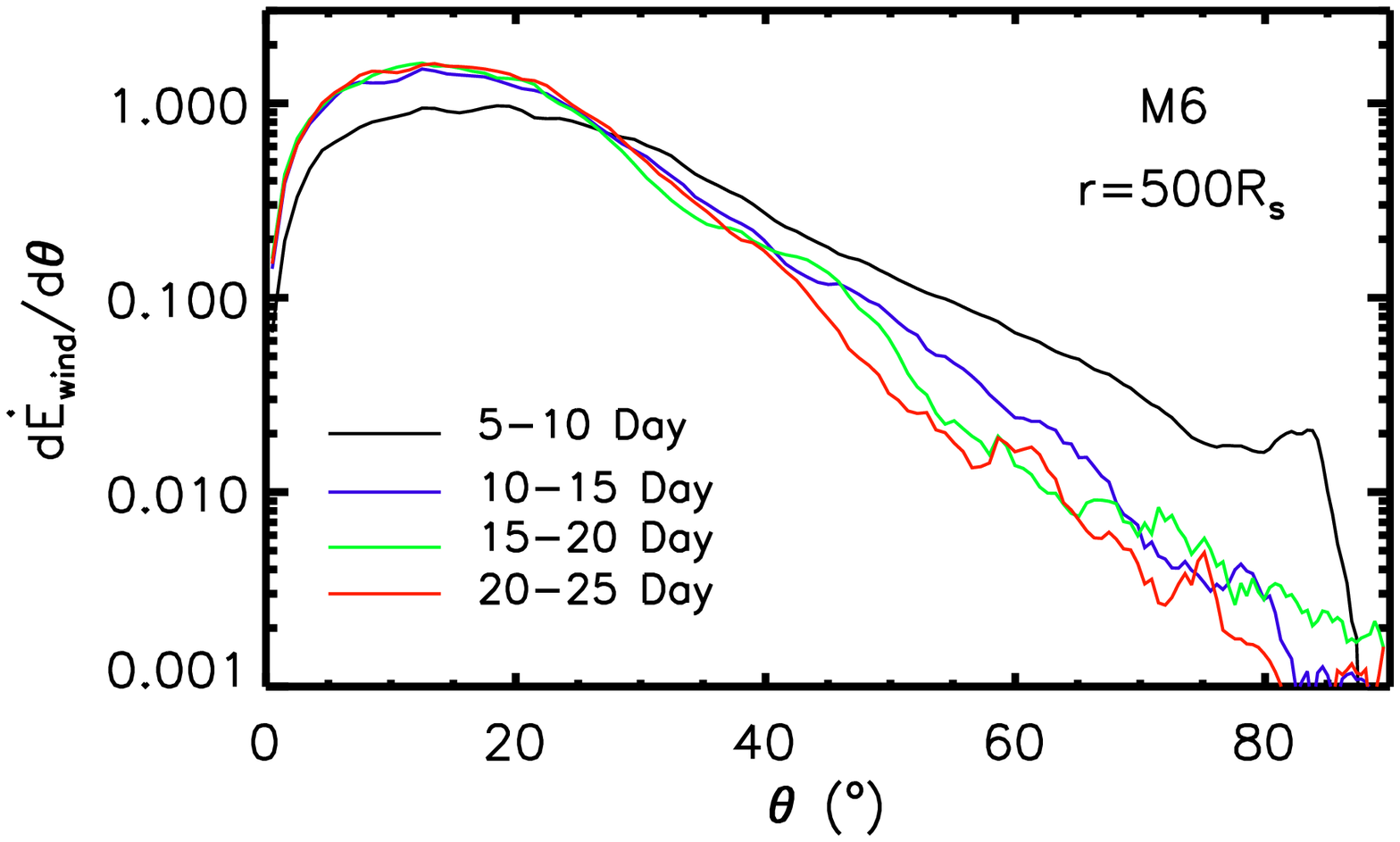}
\includegraphics[width=0.42\textwidth]{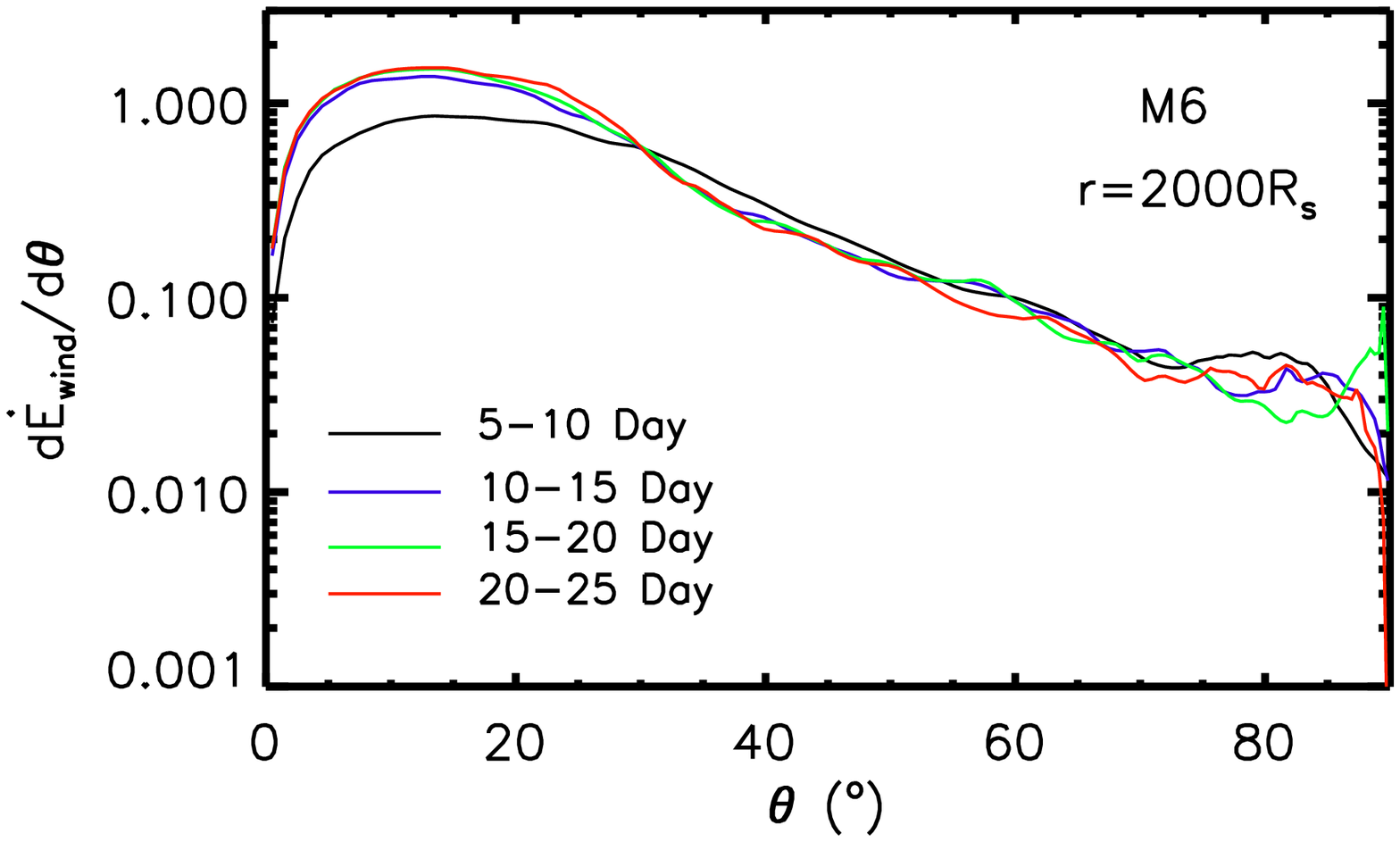}
\caption{The angular ($\theta$) distribution of the kinetic power of wind in unit of $L_{\rm Edd}$ for model M6. In order to eliminate the fluctuation, we do time-average to the kinetic power. The black line, blue line, green line and red line correspond to average period of 5-10, 10-15, 15-20 and 20-25 days, respectively. The top panel is for 500$R_s$. The bottom panel is for 2000$R_s$. }
\label{fig:powerthetaM6}
\end{center}
\end{figure}

The mass taken away by wind is only $12\%$ of the injected mass. As introduced above, the mass be accreted to the black hole is $43\%$ of the injected mass. We find that there is gas with negative Bernoulli parameter, which is just doing turbulent motions around 100$R_s$. We note that the simulation of Model M6 just covers 32 days since the peak fallback rate. The wind may have not sufficiently developed. In future, it is very necessary to run simulation with much longer physical period to study the further evolution of wind.

The angular distribution of the mass flux of wind is shown in Figure \ref{fig:moutthetaM6}. We do time-average to eliminate the fluctuation. The top panel is for 500$R_s$ and the bottom panel is for $2000R_s$. As in the case of model M7, the mass flux of wind is lowest close to the rotational axis due to the low density there.  The mass flux increases from $\theta = 0^\circ $ to $\theta = 20^\circ$. In the region $20^\circ < \theta < 90^\circ$, the mass flux of wind is roughly a constant with $\theta$ angle.

In Figure \ref{fig:vrthetaM6}, we plot the radial profile of the radial velocity of wind along several viewing angles for model M6. As in the case of Model M7, the velocity of wind is highest around the rotational axis. The velocity decreases from the rotational axis towards the midplane. The velocity around the rotational axis can be as high as 0.7$c$. The velocity at the midplane is 1 order of magnitude lower than that around the rotational axis. At small viewing angle, the velocity of wind is roughly a constant with radius. The reason is as follows. At small viewing angle, the velocity of wind is significantly higher than the local escape speed. Or in other words, the kinetic energy of wind is significantly larger than the gravitational energy, the gravity can hardly decelerate the wind.

The radial distributions of the kinetic power of wind are shown in Figure \ref{fig:powerM6}. In the region $r< 200R_s$, the kinetic power increases quickly with the radius. This is due to the fact that both the mass flux (see Figure \ref{fig:moutM6}) and the velocity (see Figure \ref{fig:vrthetaM6}) of wind increase with radius. Outside $200R_s$, the kinetic power keeps roughly a constant with radius until the cutoff. The roughly constant behaviour indicates again that the wind can hardly be decelerated by the gravity of the black hole. With the increase of time, the cutoff radius of the kinetic power increases. The kinetic power of wind can achieve $2 \times 10^{44} {\rm erg \ s^{-1}}$, which is enough to account for the radio emission in radio TDEs.

We show the angular distribution of the kinetic power of wind in Figure \ref{fig:powerthetaM6}. The power is highest in the angular region very close to the rotational axis $5^\circ < \theta < 20^\circ$. The reason is as follows. From Figure \ref{fig:vrthetaM6}, we can see that the velocity around the rotational axis is highest. The mass flux in this angular region is comparable to that in other angular region (Figure \ref{fig:moutthetaM6}). In the region $\theta > 20^\circ$, the kinetic power decreases very quickly with $\theta$. The kinetic energy flux at the midplane can be more than 2 orders of magnitude lower than that in the region $5^\circ < \theta < 20^\circ$. The reason is that in this region $\theta > 20^\circ $, the mass flux is roughly a constant with $\theta$ (see Figure \ref{fig:moutthetaM6}). However, the velocity of wind decreases very quickly with $\theta$ (see Figure \ref{fig:vrthetaM6}). The velocity at the midplane can be 1 order of magnitude lower than that around the rotational axis. The kinetic power is $\propto v^3$. Therefore, the kinetic power at the midplane is significantly low.

\section{Summary and discussions}

We study the black hole accretion and wind of circularized accretion flow in TDEs based on the radiative hydrodynamic simulations of BU22. We assume that a solar type star is disrupted. We have two models with $M_{\rm BH}=10^6M_\odot$ and $10^7M_\odot$. We also assume that the orbital pericenter of the disrupted star equals to the tidal radius. When the debris falls back, we assume that it can be very quickly circularized. An accretion flow is formed in the presence of viscosity.

The first issue we study is the relationship between the black hole accretion rate and the debris fallback rate. We find that only a part of the fallback debris can be accreted to the black hole. Specifically, for a $10^7M_\odot$ black hole, $15\%$ of the fallback debris is accreted by the black hole; for a $10^6M_\odot$ black hole, $43\%$ of the fallback debris is accreted by the black hole.

The second issue we study is the wind. We find that wind can be launched by radiation pressure in the super-Eddington accretion phase of TDEs. For a $10^7M_\odot$ black hole, more than $81\%$ of the fallback debris is taken away by wind; for a $10^6M_\odot$ black hole, $12\%$ of the fallback debris is taken away by wind. The velocity of wind decreases from the rotational axis to the midplane. Close to the rotational axis, the maximum velocity of wind can reaches 0.7$c$. At the midplane, the velocity of wind is $\sim 0.1c$. The kinetic power of wind is in the range of $(2 - 6.5) \times 10^{44} {\rm erg \ s^{-1}}$.

In model M6, the simulation just covers 32 days since the peak fallback rate. From the bottom panel of Figure \ref{fig:mdotM6}, we can see that the ratio of the black hole accretion rate to the debris fallback rate has not settled to a constant value. It is unknown, what will this value evolve further. It is very necessary to run the simulations further to cover several hundred days. In that case, a more solid conclusion about black hole accretion rate and wind can be made.

In our simulations, magnetic field is not included. It is well known that wind can be launched from an accretion flow by the magneto-centrifugal force \citep{Blandford1982}. If magnetic field is taken into account, the specific results about the properties of wind may be changed. In future, it is very necessary to study the wind by taking into account both magnetic field and the specific conditions of TDEs.

In our simulations, we use a viscous stress to transfer angular momentum. The value of $\alpha$ is set to be 0.1. \cite{Sadowski2015} found that in their super-Eddington simulations, the ratio of magnetic pressure to total pressure $p_{\rm mag}/p_{\rm tot} \sim 0.1 $. The value of $\alpha$ is defined as $\alpha = - B_r B_\phi/(4\pi p_{\rm tot})$, with $B_r$ and $B_\phi$ being the radial and toroidal components of the magnetic field. Because $B_r B_\phi/(4 \pi) \sim p_{\rm mag}$, we have $\alpha \sim p_{\rm mag}/p_{\rm tot}$. Thus, the value of alpha of a super-Eddington accretion flow can be $\sim 0.1$. Therefore, the value of alpha used in our simulations is similar to that given by magnetohydrodynamic simulations. The alpha viscosity in our work seems to be sufficient to drive accretion.

In \cite{Dai2018}, the authors find strong winds in their simulations. However, they mainly pay attention to the radiation properties of the flow. There are no detailed descriptions of the properties of winds (e.g., kinetic power, mass flux). They plot a Figure (right panel of their Figure 3) to show the velocity of winds. Their results are as follows. First, generally, they find that the velocity of wind decreases from the rotational axis towards the midplane. Second, the maximum velocity of wind is around 0.7c. Third, the minimum velocity of wind around the midplane is below 0.1c. The three properties of wind found in \cite{Dai2018} are consistent with that found in our work.

We assume that `circularization' of fallback debris is efficient. However, the efficiency of `circularization' of the fallback debris is still under debate (\cite{Kochanek1994}; \cite{Hayasaki2016}; \cite{Bonnerot2016}; \cite{Bonnerot2017}; \cite{Bonnerot2020}; \cite{Rossi2021}).
For less bound or less circularized gas, the mechanical energy (gravitational energy plus kinetic energy) is higher than that of well-circularized Keplerian flow. In this sense, the injected gas in our simulations has artificially lower mechanical energy compared to less circularized gas. It seems that gas with higher energy is much easier to produce winds. If in reality the fallback debris can be accreted by the black hole before well-circularization, we may underestimate the strength of wind by our simulations. If the circularization process of the fallback debris can be finished in a much shorter timescale compared to the accretion timescale. The debris will be first circularized and then be accreted to the black hole. The results found in our simulations should be applicable to the accretion phase. However, the winds which may be launched in the prior ‘circularization process’ need further investigations.
However, we note that for super-Eddington accretion flow, the presence of radiation pressure driven wind should be very solid, regardless of whether the flow is circularized. For completeness, it is very necessary in future to study the wind from a not fully circularized accretion flow.

\section*{Acknowledgments}
D. Bu is supported by the Natural Science Foundation of China (grants 12173065, 12133008, 12192220, 12192223) and the science research grants from the China Manned Space Project (No. CMS-CSST-2021-B02). E. Qiao is supported by the National Natural Science Foundation of China (grant 12173048) and NAOC Nebula Talents Program.  X. Yang is supported by the Natural Science Foundation of China (grant 11973018). This work made use of the High Performance Computing Resource in the Core Facility for Advanced Research Computing at Shanghai Astronomical Observatory.

\section*{Data availability}
The data underlying this article will be shared on reasonable request to the corresponding author.



\bibliographystyle{rasti}
\bibliography{ref} 








\bsp	
\label{lastpage}
\end{document}